\documentclass{IEEEtran4PSCC}

\usepackage{cite}
\usepackage{balance}

\usepackage[cmex10]{amsmath}
\usepackage{amssymb,amsthm,amsfonts}
\usepackage[hyphens]{url}
\usepackage[american]{babel}

\usepackage{graphicx}
\usepackage[caption=false,font=footnotesize]{subfig}
\usepackage{multirow}
\usepackage{algorithm,algorithmic}
\makeatletter
\renewcommand{\ALG@name}{Model}
\makeatother
\usepackage{array}
\newcolumntype{K}[1]{>{\centering\arraybackslash}p{#1}}

\usepackage{pgfplots}
\usepackage{pgfplotstable}

\pgfplotsset{compat=newest,
    /pgfplots/ybar legend/.style={
    /pgfplots/legend image code/.code={\draw[##1,/tikz/.cd,yshift=-0.25em] (0cm,0cm) rectangle (3pt,0.8em);},
   },
   }

\usepackage{color}
\usepackage{enumerate}

\definecolor{myred}{RGB}{202,0,32}
\definecolor{myorange}{RGB}{244,165,130}
\definecolor{myviolet}{RGB}{194,165,207}
\definecolor{mycyan}{RGB}{146,197,222}
\definecolor{myblue}{RGB}{5,113,176}
\definecolor{mygreen}{RGB}{127,191,123}
\definecolor{mytile}{RGB}{27,120,55}

\usepackage{tikz}

\newtheorem{lemma}{Lemma}[section]

\newtheorem{prop}[lemma]{Proposition}

\newtheorem{thm}[lemma]{Theorem}

\theoremstyle{remark}
\theoremstyle{remark}
\theoremstyle{definition}

\newcommand{\set}[1]{\left\{#1\right\}}
\newcommand{\abs}[1]{\left|#1\right|}

\DeclareMathOperator{\diag}{diag}

\newcommand{\R}{\mathbb R}

\newcommand{\N}{\mathbb N}

\newcommand{\ul}[1]{\underline{#1}}
\newcommand{\ol}[1]{\overline{#1}}

\newcommand{\goesto}{\rightarrow}

\newcommand{\calE}{\mathcal{E}}

\newcommand{\calG}{\mathcal{G}}

\newcommand{\calN}{\mathcal{N}}

\newcommand{\calP}{\mathcal{P}}

\newcommand{\calR}{\mathcal{R}}

\newcommand{\calT}{\mathcal{T}}

\usepackage{color}
\usepackage{comment}
\usepackage{ifthen}
\newboolean{showcomments}
\setboolean{showcomments}{true}
\newcommand{\lguo}[1]{\ifthenelse{\boolean{showcomments}}
{ \textcolor{orange}{(Daniel says:  #1)}}{}}
\newcommand{\slow}[1]{\ifthenelse{\boolean{showcomments}}
{ \textcolor{blue}{(Steven says:  #1)}}{}}
\newcommand{\cliang}[1]{\ifthenelse{\boolean{showcomments}}
{ \textcolor{myred}{(Chen says:  #1)}}{}}
\newcommand{\alex}[1]{\ifthenelse{\boolean{showcomments}}
{ \textcolor{green}{(Alex says:  #1)}}{}}

\makeatletter
\let\old@ps@headings\ps@headings
\let\old@ps@IEEEtitlepagestyle\ps@IEEEtitlepagestyle
\def\psccfooter#1{%
    \def\ps@headings{%
        \old@ps@headings%
        \def\@oddfoot{\strut\hfill#1\hfill\strut}%
        \def\@evenfoot{\strut\hfill#1\hfill\strut}%
    }%
    \def\ps@IEEEtitlepagestyle{%
        \old@ps@IEEEtitlepagestyle%
        \def\@oddfoot{\strut\hfill#1\hfill\strut}%
        \def\@evenfoot{\strut\hfill#1\hfill\strut}%
    }%
    \ps@headings%
}
\makeatother

\psccfooter{%
        \parbox{\textwidth}{\hrulefill \\ \small{21st Power Systems Computation Conference} \hfill \begin{minipage}{0.2\textwidth}\centering \vspace*{4pt} \includegraphics[scale=0.06]{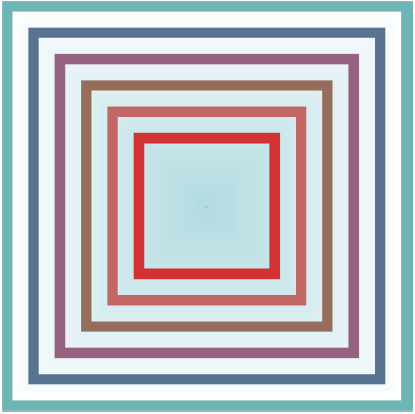}\\\small{PSCC 2020} \end{minipage} \hfill \small{Porto, Portugal --- June 29 -- July 3, 2020}}%
}

\begin{document}

\title{An Integrated Approach for Failure Mitigation \& Localization in Power Systems}


 \author{\IEEEauthorblockN{Chen Liang\IEEEauthorrefmark{1},
 Linqi Guo\IEEEauthorrefmark{1},
 Alessandro Zocca\IEEEauthorrefmark{2}, 
 Shuyue Yu\IEEEauthorrefmark{1},
 Steven H. Low\IEEEauthorrefmark{1} and
 Adam Wierman\IEEEauthorrefmark{1}}
 \IEEEauthorblockA{\IEEEauthorrefmark{1} Department of Computing and Mathematical Sciences, California Institute of Technology, Pasadena, USA \\  \{cliang2, lguo, syu5, slow, adamw\}@caltech.edu}
 \IEEEauthorblockA{\IEEEauthorrefmark{2} Department of Mathematics, Vrije Universiteit, Amsterdam, Netherlands \\ a.zocca@vu.nl}
 }

\maketitle

\begin{abstract}
    The transmission grid is often comprised of several control areas that are connected by multiple tie lines in a mesh structure for reliability.  It is also well-known that line failures can propagate non-locally and redundancy can exacerbate cascading. In this paper, we propose an integrated approach to grid reliability that (i) judiciously switches off a small number of tie lines so that the control areas are connected in a tree structure; and (ii) leverages a unified frequency control paradigm to provide congestion management in real time. Even though the proposed topology reduces redundancy, the integration of tree structure at regional level and real-time congestion management can provide stronger guarantees on failure localization and mitigation. We illustrate our approach on the IEEE 39-bus network and evaluate its performance on the IEEE 118-bus, 179-bus, 200-bus and 240-bus networks with various network congestion conditions. Simulations show that, compared with the traditional approach, our approach not only prevents load shedding in more failure scenarios, but also incurs smaller amounts of load loss in scenarios where load shedding is inevitable. Moreover, generators under our approach adjust their operations more actively and efficiently in a local manner.
\end{abstract}


\thanksto{\noindent 
This work has been supported by Resnick Fellowship, Linde Institute Research Award, NWO Rubicon grant 680.50.1529, NSF through awards ECCS 1619352, CNS 1545096, CCF 1637598, ECCS 1739355, CNS 1518941, CPS 154471,  ARPA-E through award DE-AR0000699 (NODES), and DTRA through award HDTRA 1-15-1-0003.}

\section{Introduction}
Reliability is critical in power systems.
Tremendous efforts from both the industry and academia have been made to analyze cascading failures. Current industry practice is typically simulation-based,  where contingencies are studied using extensive simulations \cite{vaiman2012risk}. Such approaches are often limited by computational power. 

To provide tractable analysis, pure topological models have been proposed, where failures propagate locally to neighboring components with high probability \cite{brummitt2003cascade,kong2010failure,crucitti2004topological}. However, these epidemic models are not realistic as non-local failure propagation is observed in both real-world and simulated cascade data \cite{hines2017cascading}. More realistic models use linearized DC power flow to characterize power redistribution after transmission line failures \cite{carreras2002critical,bienstock2010n-k,bienstock2011control,soltan2015analysis}. 
These DC models indeed exhibit both local and non-local propagation of failures. See \cite{guo2017critical} for an extensive list of cascading failure models.
It is observed in \cite{soltan2015analysis} that successive failures can be quite far away from initial failures under the DC model, which aligns with real cascade data.

Such non-local failure propagation comes from the interconnectivity of the power network.
The transmission grid is usually comprised of several control areas, which are operated relatively independently with prescribed power exchanges determined by economic dispatch and maintained by automatic generation control (AGC) \cite{bergen2000power}.  
Traditionally, control areas are interconnected by multiple tie lines in a \emph{mesh} structure to provide multiple alternative routes for power, as redundancy often improves reliability\cite{machowski2011power}. Surprisingly, it is shown in \cite{guo2018failure} that \emph{tree partitioning} of the grid insulates the impact of failures and precisely captures the boundaries of failure propagation.  This means that, while providing redundancy, multiple tie lines also exacerbate non-local failure propagation. 

Failure models based on DC power flow often assume power injections remain unchanged after a failure as long as the network remains connected. This is a reasonable assumption under the traditional frequency control dynamics that operate at a fast timescale (see Section~\ref{section:pre}), although this connection has not been widely mentioned in the literature. An important feature of our work is that we explicitly model the interaction between frequency control dynamics at a fast timescale and the DC power flow at a slow timescale in the cascading process, and leverage this interaction for failure mitigation.

In this paper, we propose an integrated approach to grid reliability consisting of two main components: \emph{topology design} and \emph{real-time response}. For topology design, we propose to judiciously switch off a small number of tie lines so that the control areas are interconnected in a tree structure. At real-time, we leverage the recently proposed unified frequency control \cite{mallada2017optimal} to provide congestion management at a fast timescale. Even though the proposed approach reduces redundancy in network topology, the integrated design of tree-connected control areas and real-time congestion management provides stronger guarantees on failure localization and mitigation, leading to a higher overall reliability.
This new framework builds on our earlier work \cite{guo2018failure,guo2019less}.  Here, we extend the approach and evaluate its performance in IEEE test networks.

We show in Tables \ref{table:stat1} and \ref{table:stat2} of Section \ref{section:case_studies} using simulations over the IEEE 118-bus, 179-bus, 200-bus and 240-bus networks that our proposed approach not only prevents load shedding in more failure scenarios, but also incurs smaller amounts of load loss in scenarios where load shedding is inevitable. Moreover, generators under our approach respond more actively and efficiently to terminate cascading failures with only local adjustments.

The rest of the paper is organized as follows. We review in Section~\ref{section:pre} the DC failure model and propose an integrated failure model that incorporates the effect of frequency control on failure propagation. Our proposed approach is presented in Section~\ref{section:philosophy} with implementation details on topology design and real-time response. We then demonstrate the theoretical guarantees and benefits of our approach in Section~\ref{section:remark}. Lastly, we illustrate and evaluate our approach over the IEEE test networks in Sections~\ref{section:example} and \ref{section:case_studies}.
All proofs are omitted and can be found in a detailed 3-part paper under preparation.

\section{Failure Models} \label{section:pre}
To begin, we review a widely used DC failure propagation model, and then propose an integrated model that extends the DC model to incorporate frequency control dynamics.

\subsection{DC Failure Model}
Power grids are usually modeled by a set of non-linear and non-convex AC power flow equations \cite{bergen2000power}. It is, however, less efficient for large-scale power networks. In this paper, we adopt the linearized DC power flow for tractability, which is widely used in contingency analysis \cite{carreras2002critical,bienstock2010n-k,bienstock2011control,soltan2015analysis,ba2019computing}.
In particular, we represent the power transmission network by a directed graph $\calG=(\calN, \calE)$, where $\calN=\{1,2,\dots,n\}$ and $\calE=\{e_1,\dots,e_m\}$ are the sets of buses and transmission lines, respectively. The terms bus/node and line/edge are used interchangeably in this paper. An edge $e$ in $\calE$ between nodes $i$ and $j$ is denoted as either $e$ or $(i,j)$, and an arbitrary direction is assigned to each edge. We assume lines are purely reactive, and each line $e$ is characterized by its susceptance $b_e$.

The cascading process is described in stages indexed by $k\in\N$. The topology at stage $k$ is denoted as $\calG(k) = (\calN,\calE(k))$ and $\calG(0)$ denotes the original network. Given the power injections and phase angles $p(k),\theta(k) \in \R^n$, the line flows $f(k) \in \R^{m(k)}$ are the solution to the following DC power flow equations:
\begin{subequations}\label{eqn:dc_model}
\begin{IEEEeqnarray}{rCl}
	p(k)&=&C(k)f(k) ,\label{eqn:flow_conservation}\\
	f(k)&=&B(k)C(k)^T\theta(k), \label{eqn:kirchhoff}
\end{IEEEeqnarray}
\end{subequations}
where $B(k):=\diag\left( b_1,\dots,b_{m(k)} \right)$ is the diagonal susceptance matrix, and $C(k)\in\R^{n\times m(k)}$ is the  node-edge incidence matrix. Rows and columns of $C(k)$ correspond to nodes and edges. For every edge $e=(i,j) \in \calE(k)$, we set $C_{i, e}(k) = 1$ and $C_{j, e}(k)=-1$, while all other entries are set to zero.

In order for the linear system \eqref{eqn:dc_model} to have a solution, the
 power must be balanced on each island of the network, i.e., $\sum_{i\in\calN_l} p_i(k)=0$ for each connected component $\calN_l$ of $\calG(k)$. 
Under this condition, line flows are uniquely determined by
$$f(k)=B(k)C(k)^T \left( C(k)B(k)C(k)^T \right)^\dagger p(k),$$
where $(\cdot)^\dagger$ denotes the Moore-Penrose inverse.

We now formally describe the DC failure model. The cascade starts from an initial set of line failures and propagates in stages. At each stage $k\in\N^+$, assume a set $E(k) \subset \calE(k-1)$ of lines fail. The power injections $p(k)$ first adjust based on certain balancing rule $\calR$ for each island, then the line flows redistribute over the new topology $\calG(k) = (\calN, \calE(k))$, where $\calE(k) = \calE(k-1) \setminus E(k)$. We adopt the following deterministic outage rule: every line $e$ with power flow exceeding its line limit $\pi_e$ is tripped at the next stage, i.e., $E(k+1) =  \{e: \abs{f_e(k)} > \pi_e, e\in\calE(k)\}$. If all lines are within their limits ($E(k+1)=\emptyset$), the cascade is terminated; otherwise the process repeats for stage $k+1$.

The evolution of the cascading failure critically depends on the power balancing rule $\calR$. A commonly used rule $\calR_c$ in the cascading failure literature can be described as follows \cite{soltan2015analysis}: (i) If the network remains connected after the failure $E(k)$, then $p(k)$ remains the same as previous stage $p(k-1)$; otherwise (ii) all the nodes proportionally adjust their injections to compensate for the imbalance on each island.
As explained in the next subsection, this rule can be interpreted as a special case of the integrated failure model that we now present.

\subsection{Integrated Failure Model}
From this subsection on, we consider the failure dynamics for a single stage and drop the index $k$ for presentation clarity. We use superscript $(\cdot)^0$ to denote the pre-contingency nominal steady-state values, and symbols followed by $(t)$ to indicate the dynamic process.

Assume the pre-contingency grid $\calG^0=(\calN,\calE^0)$ operates at a nominal steady-state $(f^0, p^0, \theta^0)$ which satisfies the DC power flow equation \eqref{eqn:dc_model}, i.e. $f^0 = B^0 C^{0T} (C^0B^0C^{0T})^\dagger p^0$. Given a set $E$ of line failures, we use $\calE := \calE^0 \setminus E$ to denote the surviving lines. Let $B, C$ denote the susceptance and incidence matrices for the surviving network $\calG = (\calN, \calE)$. The dynamics for bus frequency \emph{deviations} $\omega(t) \in\R^{\abs{\calN}}$ and line flows $f(t) \in \R^{\abs{\calE}}$ over the remaining lines can be described by the following linear swing and power flow equations~\cite{bergen2000power}:%
\begin{subequations}\label{eqn:frequency_con}
\begin{IEEEeqnarray}{rCl}
  M \dot{\omega}&=& p^0 - d(t) - D \omega(t) - C f(t), \label{eqn:swing_dynamics}\\
  \dot{f} &=& BC^T \omega(t),\label{eqn:network_flow_dynamics}
\end{IEEEeqnarray}
\end{subequations}
where $M \in \R^{\abs{\calN} \times \abs{\calN}}$ is the diagonal matrix containing information about the system inertia, $d(t) \in \R^{\abs{\calN}}$ is the \emph{deviation} on controllable power injections (such as droop control, generator ramping in
response of power imbalance, and load-side participation),
and $D \omega(t) \in \R^{\abs{\calN}}$ denotes the \emph{deviation} on system damping as well as load dynamics. The initial values for dynamics \eqref{eqn:frequency_con} are $\omega(0) = 0$ and $f(0) = f^0_{\calE}$. Note here that the pre-contingency injection $p^0$ should be interpreted as the sum of nominal values of generator injection, controllable injection, load, and system damping. As such, $d(t)$ represents the system response in terms of the controllable load deviation after a failure event.

A state $x^*:=(d^*, \omega^*,f^*) \in \R^{\abs{\calN}} \times \R^{\abs{\calN}} \times \R^{\abs{\calE}}$ is said to be an \emph{equilibrium}\footnote{We do not explicitly model the phase angle dynamics $\dot{\theta}=\omega$ and, hence, do not enforce zero frequency deviation at equilibrium in our setup. This allows \eqref{eqn:frequency_con} to model both primary and secondary frequency control, demonstrating their close connections in a common framework. See \cite{guo2019tps3} for more discussions.}
if the right-hand sides of \eqref{eqn:frequency_con} are zero at $x^*$. 
It is clear that, at equilibrium, the post-contingency line flows $f^*$ satisfy the DC power flow equations with \emph{post-contingency} injections $p^*=p^0-d^*-D\omega^*$ over \emph{post-contingency} network $\calG$. This fact suggests that the power balancing rule mentioned in the previous subsection can be interpreted as a result of the linear frequency dynamics. Indeed, we show that when classical droop control is adopted, i.e. $d_j(t)=\alpha_j \omega_j(t)$, the balancing rule $\calR_c$ is recovered.

As shown in \cite{zhao2016unified}, the equilibrium of \eqref{eqn:frequency_con} under droop control is the 
optimal solution of the following optimization\footnote{For simplicity, we assume the constraints on control actions $d$ are inactive.}:%
\begin{subequations}\label{eqn:droop_olc}
\begin{IEEEeqnarray}{ll}
\min_{\omega, d, f, \theta} \quad & \sum_{j\in\calN}\frac{d_j^2}{2\alpha_j}+\frac{D_j w_j^2}{2}\label{eqn:droop_obj}\\
 \hspace{.2cm}\text{s.t.}& p^0 - d - D\omega = Cf,\label{eqn:droop_balance}\\
	& f - BC^T  \theta = 0.\label{eqn:droop_branch}
\end{IEEEeqnarray}
\end{subequations}
If the grid becomes disconnected with several islands $\{\calN_1, \dots, \calN_l\}$, 
then the optimal solution of \eqref{eqn:droop_olc} is:
\begin{subequations}\label{eqn:droop_opt}
\begin{IEEEeqnarray}{rCll}
d_j^* &=& \frac{\alpha_j}{\sum_{i\in \calN_k} (\alpha_i+D_i)} \sum_{i\in \calN_k} p_i^0 ,& \text{\quad for $j\in \calN_k$}, \\
D_j \omega_j^* &=& \frac{D_j}{\sum_{i\in \calN_k} (\alpha_i+D_i)} \sum_{i\in \calN_k} p_i^0, & \text{\quad for $j\in \calN_k$}.
\end{IEEEeqnarray}
\end{subequations}
Thus, in the equilibrium state 
the power injections adjust linearly in the power imbalance $\sum_{i\in \calN_k} p_i^0$ on each island $\calN_k$ of the post-contingency network $\calG$, precisely as prescribed by the power balancing rule $\calR_c$.

We now describe our integrated failure model, which extends DC power flow at a slow timescale by incorporating frequency dynamics at a fast timescale.
After line failures, instead of solving DC power flow equations with balancing rule $\calR$, the system evolves according to the frequency dynamics \eqref{eqn:frequency_con} and converges to an equilibrium. As before, a line trips in the next round if and only if its \emph{steady-state} flow exceeds its capacity, while an overload during transient does not cause a line failure. This is a reasonable assumption, as line outages normally require time for thermal accumulation.

Compared with DC model, there are many benefits of this integrated failure model. First, it provides a clear explanation of the power balancing rules already introduced in the literature. Indeed, the validity of various balancing rules can be justified in a similar manner from a frequency dynamic perspective. Second, the equilibrium of \eqref{eqn:frequency_con} can usually be efficiently obtained from optimization problems like \eqref{eqn:droop_olc}. Tractable analysis can thus be performed without simulating transient dynamics. More importantly, the integrated failure model offers a systematic method to analyze network evolution under various control actions, allowing us to \emph{reverse engineer} the controller $d(t)$ in order to  find a potentially better system response to failures. Our proposal of integrating the unified controller, to be presented in next section, in the context of failure mitigation is an example of how we can leverage this model to improve the system reliability in achieving desirable control properties.

\section{An Integrated Approach to Failure Mitigation} \label{section:philosophy}
In this section, we first give a high-level description of our proposed approach and then provide technical and implementation details on the topology design and real-time response.

\subsection{Overview}
Our proposal to improve grid reliability consists of two main components: 
\emph{topology design} and \emph{real-time response}.  It aims to achieve the following desirable properties:
\begin{itemize}
    \item \emph{Optimal mitigation}: Cascading failures should be stopped, and system adjustments for generations and loads (including load shedding) should be minimized.

    \item \emph{Local impact}:  Disturbances in a control area should not impact other areas if at all possible. Control areas can thus be operated relatively independently.
    
    \item \emph{Autonomous response}: The control actions should be implemented in real-time in an autonomous and distributed manner. Current approaches to failure mitigation often involves human in the loop, rendering it slower, less optimal and possibly more error-prone.
\end{itemize}

For the topology design, we propose a tree structure at the control area level, contrary to the conventional design where areas are connected in a mesh structure through multiple tie lines for grid reliability. 
At real-time, we adopt a unified controller for frequency regulation that also manages congestion at fast timescale. A distributed detection algorithm is implemented in parallel to assess the severity of failures and adjust the controller to stabilize the grid when necessary. 

\subsection{Topology Design}
Redundancy has been the key mechanism for grid reliability, e.g., the $N-1$ security standard \cite{bergen2000power,bienstock2007integer,hines2007controlling,albert2004structural}. Different control areas in current transmission grids are thus mesh-connected by multiple tie lines in order to provide multiple alternative routes for power to flow through. 

It has been shown recently that such a redundancy-based design allows the impact of failures to propagate more broadly, while tree partitioning of the network guarantees control area independence, i.e., line failures are constrained within their own control area \cite{guo2018failure}. 
We thus propose to judiciously switch off a small number of tie lines so that the resulting control areas are connected in a tree structure, aiming to improve grid reliability through better failure localization.

More specifically, consider a grid $\calG=(\calN,\calE)$ with control areas described by a partition $\calP:=\{\calN_1, \dots, \calN_l\}$, where $\calN_i \cap \calN_j=\emptyset$ for $i\neq j$ and $\bigcup_{i=1}^l \calN_i = \calN$. We denote $\calT(\calP):=\set{(s,t)\in\calE| s\in \calN_i, t\in\calN_j, i\neq j}$ as the set of all tie lines connecting different areas.
The reduced graph $\calG_\calP(\calE)$ under partition $\calP$ is a graph obtained from $\calG$ by collapsing each area $\calN_i$ into a ``super node'' and adding an edge between super nodes $\calN_i$ and $\calN_j$ for each tie line connecting them. As mentioned earlier, the redundancy-based design usually leads to a non-simple (i.e., there may be multiple lines between two super nodes)  or cyclic reduced graph.
Our method aims to select a subset of tie lines $T\subset \calT(\calP)$ to switch off, such that the control areas of the remaining network are connected in a tree topology, i.e. the reduced graph $\calG_\calP(\calE \setminus T)$ is a tree. This implies that $\abs{\calT(\calP)}-l+1$ tie lines will be switched off.

Similarly to line failures, tie line switching actions change the system operating point as power flows redistribute in the new network topology.
Let $p$ denote the average injections for topology design purpose.
We choose the set $T$ of candidate lines to minimize 
\emph{network congestion level} $\gamma(T)$ defined as:
$$
\gamma(T) = \max_{e \in \calE \setminus T} \abs{f_e(T)} / \pi_e,
$$
where $f_e(T)$ is the line flow on line $e$ after the lines in $T$ are switched off.
We are therefore interested in solving the optimization problem:
\begin{subequations}\label{eqn:opt_sw}
\begin{IEEEeqnarray}{ll}
\min_{T\subset \calT(\calP)} \quad &  \gamma(T)  \\
 \hspace{.3cm}\text{s.t.}& \text{$\calG_\calP(\calE \setminus T)$ is a tree.}
\end{IEEEeqnarray}
\end{subequations}

The complexity of the optimization \eqref{eqn:opt_sw} originates from finding all possible subsets $T$ of $\calT(\calP)$ to switch off. 
Solving the above optimization problem often becomes intractable for large-scale power grids. Nevertheless, such switching actions are only implemented occasionally, rather than continuously in real time. An approximate but faster algorithm is proposed in \cite{zocca2019or} where tree-connected areas are created by recursively splitting the existing ones, yielding very good results for most application scenarios.

We remark that it is not guaranteed that $\gamma(T^*) \leq 1$, where $T^*$ is an optimal selection, implying that some transmission lines may become overloaded after switching actions. This may be alleviated if one has the flexibility to design the control areas of the grid. We refer interested readers to \cite{zocca2019or} for optimal partitioning of the grid using network modularity clustering algorithms. However, $\gamma(T^*) < 1$ indeed holds for most practical scenarios simulated in \cite{zocca2019or}, especially when the original grid is not heavily congested.

\subsection{Real-time Response}
Once a tree-connected control area structure is created, the unified controller (UC) is implemented as a frequency regulation method to autonomously respond to disturbances such as loss of generation/load and line failures. 
The closed-loop dynamics of UC are more elaborate than \eqref{eqn:frequency_con}; see \cite{zhao2016unified, mallada2017optimal} for details. It is shown there that, under mild conditions, the closed-loop equilibrium under UC is globally asymptotically stable.  Moreover it is the optimal solution of the following optimization:
\begin{subequations}\label{eqn:uc_olc}
\begin{IEEEeqnarray}{ll}
\min_{f, d, \theta} \quad& \sum_{j\in\calN} \frac{d_j^2}{2\alpha_j} \label{eqn:uc_obj}\\
\hspace{.1cm}\text{s.t.} & p^0 - d - C f = 0, \label{eqn:uc_balance}\\
& f = BC^T \theta,  \label{eqn:dcflow}\\
& E_{T} C f = E_{T} C^0f^0,\label{eqn:uc_ace}\\
& \ul{f}_{e}\le f_{e}\le \ol{f}_{e}, \quad e\in \calE, \label{eqn:line_limit}\\
& \ul{d}_{j}\le d_j\le \ol{d}_j, \quad j \in\calN, \label{eqn:control_limit}
\end{IEEEeqnarray}
\end{subequations}
where \eqref{eqn:uc_balance} and \eqref{eqn:dcflow} describe the DC power flow equations, \eqref{eqn:uc_ace} ensures zero area control error, \eqref{eqn:line_limit} enforces line flow limits, and \eqref{eqn:control_limit} enforces control limits.
Matrix $E_T$ describes the control areas $\calP=\{\calN_1, \dots, \calN_l\}$, namely $E_T\in\{0,1\}^{l \times \abs{\calN}}$ is defined as $E_{T(i,j)}=1$ if node $j\in \calN_i$ and $E_{T(i,j)}=0$ otherwise. Therefore, the $i$-th row of the constraint $E_TCf= E_T C^0f^0$ ensures that the net total power flow on tie lines connected to area $\calN_i$ is restored to the pre-contingency value\footnote{{If a control area is disconnected from the grid, the corresponding row of \eqref{eqn:uc_ace} will be lifted to balance the injections.}}. 

Explaining the design of UC (which prescribes the dynamics of controllable power injection $d(t)$) is beyond the scope of this paper; interested readers are referred to \cite{zhao2016unified}. Nevertheless, it should be noted that the controller is derived from a variant of primal-dual algorithm to solve \eqref{eqn:uc_olc}, where the primal updates are carried out by the network dynamics itself, while the dual variables $\lambda$ are updated with only local communications. Under mild conditions, UC converges to the optimal solution of \eqref{eqn:uc_olc}, provided that the optimization is feasible.

If the disturbances are small, \eqref{eqn:uc_olc} is likely to be feasible and UC is then guaranteed to drive the network to an equilibrium, where the nominal frequency and the inter-area flows are restored (zero area control error), and more importantly the line limits are enforced. Such failures can thus be properly mitigated, leveraging the congestion management of UC.

However, when a \emph{severe} disturbance occurs, \eqref{eqn:uc_olc} may no longer be feasible, which implies that the controller is not capable of achieving all its control goals. If such failures happen, UC is no longer stable and can potentially lead to a large scale outage. It is thus crucial to promptly identify such severe failures and respond accordingly. To do so, we propose a distributed algorithm to assess the severity of failures in real-time, and a proactive procedure to adjust the controller to stabilize the system if necessary; see \cite{guo2019less} for more details.

\subsubsection{Severe Failure Detection}
Severe failures can be detected by monitoring the dual variables of UC in real-time. As shown in the following proposition, after a severe failure at least one dual variable grows arbitrarily large during the transient phase and the system never reaches an equilibrium point. Therefore, a warning can be raised whenever a dual variable exceeds a predefined threshold.
\begin{prop} \label{prop:dual_to_infinity}
If the optimization problem \eqref{eqn:uc_olc} is infeasible and a primal-dual update algorithm is implemented to solve \eqref{eqn:uc_olc}, then there exists a dual variable $\lambda_i$ such that
$$
\limsup_{t\goesto\infty}\abs{\lambda_i(t)} = \infty.
$$
\end{prop}
This approach is guaranteed to detect all severe failures, but it may yield false alarms as some dual variables can possibly become very large during transient even when \eqref{eqn:uc_olc} is feasible. There is thus a trade-off between the speed and accuracy of detection: a larger threshold would reduce the false alarm rate, but also requires longer time to detect severe failures. Therefore, the threshold should be selected carefully based on system parameters and application scenarios.

\subsubsection{Response to Severe Failures}
When severe failures happen, it is impossible for UC to enforce all the constraints in \eqref{eqn:uc_olc}.  
To alleviate the impact of such failures and stabilize the system, we propose to adjust the controller by proactively lifting constraints in \eqref{eqn:uc_olc} until it becomes feasible again. 

However, not all constraints can be relaxed, as some of them are essential to system stability. In particular, \eqref{eqn:uc_balance} and \eqref{eqn:dcflow} represents the physics of power flow, while \eqref{eqn:line_limit} ensures post-contingency flows are safe. The other two constraints can be lifted without compromising stability as follows.

First, some of the zero area control error constraints $E_TCf=E_TC^0f^0$ can be lifted by setting the corresponding dual variables to $0$. This action allows more control areas to adjust their operations in the mitigation of the failure.
This may be undesirable as it is counter to our goal of failure localization. 
Second, the range $[\ul{d}_j, \ol{d}_j]$ for the controllable injections can be relaxed by enlarging its width. This action usually corresponds to shedding loads, which is clearly undesirable.

By relaxing some rows of the $E_TCf=E_TC^0f^0$ constraints and allowing controlled load shedding, problem \eqref{eqn:uc_olc} can always be made feasible with all line flows satisfying their limits (more details below).  
Hence any severe failure can be terminated and will not cascade out of control, though 
with possible degradation on localization performance and potential loss of load. Therefore, such a constraint lifting procedure should be implemented carefully to prioritize the desired objectives.

\section{Discussion}\label{section:remark}
In this section, we discuss our proposed approach and demonstrate how it achieves all our design goals.

\subsection{Guaranteed Mitigation Performance}
Our proposed approach provably terminates line failures and optimally drives the system to a desired equilibrium. For these guarantees the unified frequency control is of key importance.
Assume the pre-contingency flows are within safe limits.
It can be shown that the relaxed optimization problem in which \eqref{eqn:uc_ace} is lifted and \eqref{eqn:control_limit} is enlarged by allowing all loads to be shedded, yields a trivial feasible point by reducing all generations and loads to zero.
Therefore, regardless of whether the failure is severe or not, we can always relax \eqref{eqn:uc_olc} progressively to a feasible optimization, at the cost of more affected areas and/or potential load loss. Nevertheless, given the global stability of UC, the system is guaranteed to converge to an equilibrium such that post-contingency flows are below line limits, i.e., successive failures are prevented. 

The objective \eqref{eqn:uc_obj} should be interpreted as the penalty for the control actions $d$. The optimal solution $d^*$ is exactly the adjustment of power injections at the equilibrium under the unified frequency control in response to a failure. Therefore, UC manages to mitigate the failure optimally in terms of minimizing the injection adjustments. We remark that the control actions can be further separated into generations and loads $d=d^G-d^L$, and the objective \eqref{eqn:uc_obj} can be designed to implement different priorities of generation/load adjustments.  
For instance, if the objective \eqref{eqn:uc_obj} is modified to $\sum_{j\in\calN} \frac{(d_j^G)^2}{2\alpha_j^G} + \frac{(d_j^L)^2}{2\alpha_j^L}$, then smaller $\alpha^L_j$'s can be used to prefer generation adjustments over load adjustments.
%


\subsection{Guaranteed Localization Performance}
The proposed approach is guaranteed to terminate failures with only local impact at equilibrium if the failure is not severe. The tree structure plays a crucial role in this aspect.
To demonstrate our localization result, we clarify the meaning of ``local impact" through the notion of \emph{associated areas}. We say an area $\calN_k$ is associated with failures $E$ if there exists an edge $e=(i,j)\in E$ such that either $i\in\calN_k$ or $j\in\calN_k$. Moreover, we only study the localization when the system converges to an equilibrium, while the non-local response during the transient stage is ignored.

The following theorem characterizes the localization performance of our proposed approach: after a non-severe failure the system converges to an equilibrium where the failure is mitigated with unchanged injections for non-associated areas. This guarantee extends the results in \cite{guo2018failure} in two ways: (i) both bridge\footnote{A bridge is a cut edge whose removal disconnects the network.} and non-bridge failures are localized within associated areas; (ii) failures are terminated so that successive failures with potential global impact are prevented.
\begin{thm}\label{thm:local}
Assume the control areas are tree-connected. Given a set $E$ of failures, if \eqref{eqn:uc_olc} is feasible, then $d_j^*=0$ for all $j\in\calN_k$ if $\calN_k$ is not associated with $E$.
\end{thm}
This strong localization guarantee stems from the combination of tree-connected control areas and unified frequency control. Without the tree structure, UC would only guarantee the overall power imbalance is restored in each control area, while the individual operating conditions may already vary.

\subsection{Other Benefits}
Another appealing feature of our proposed control is that it can be implemented in an autonomous and distributed fashion, allowing for real-time response to failures. The problem of how to optimally shed load to mitigate failures has already been considered in the literature e.g. \cite{bienstock2011control}. The authors solve a similar optimization problem but using a central controller that requires global information and failure information. In contrast, our approach operates as a closed-loop distributed controller and drives the system to a desired equilibrium autonomously. 

Our method has also another economic benefit. In practice, $N-1$ preventive SCOPF is solved to ensure system reliability. Our approach, however, can serve as a corrective method to mitigate non-severe failures. Thus SCOPF only need consider those severe failures, potentially leading to significant savings.

\section{An Illustrative Example}\label{section:example}
In this section, we illustrate the dynamic response of our approach for the IEEE 39-bus network with parameters adapted from \cite{zhao2016unified}. It consists of two control areas, which are connected by three tie lines, namely $(1,2)$, $(2,3)$ and $(26,27)$.
\begin{figure}[t]
\centering
\includegraphics[width=.4\textwidth]{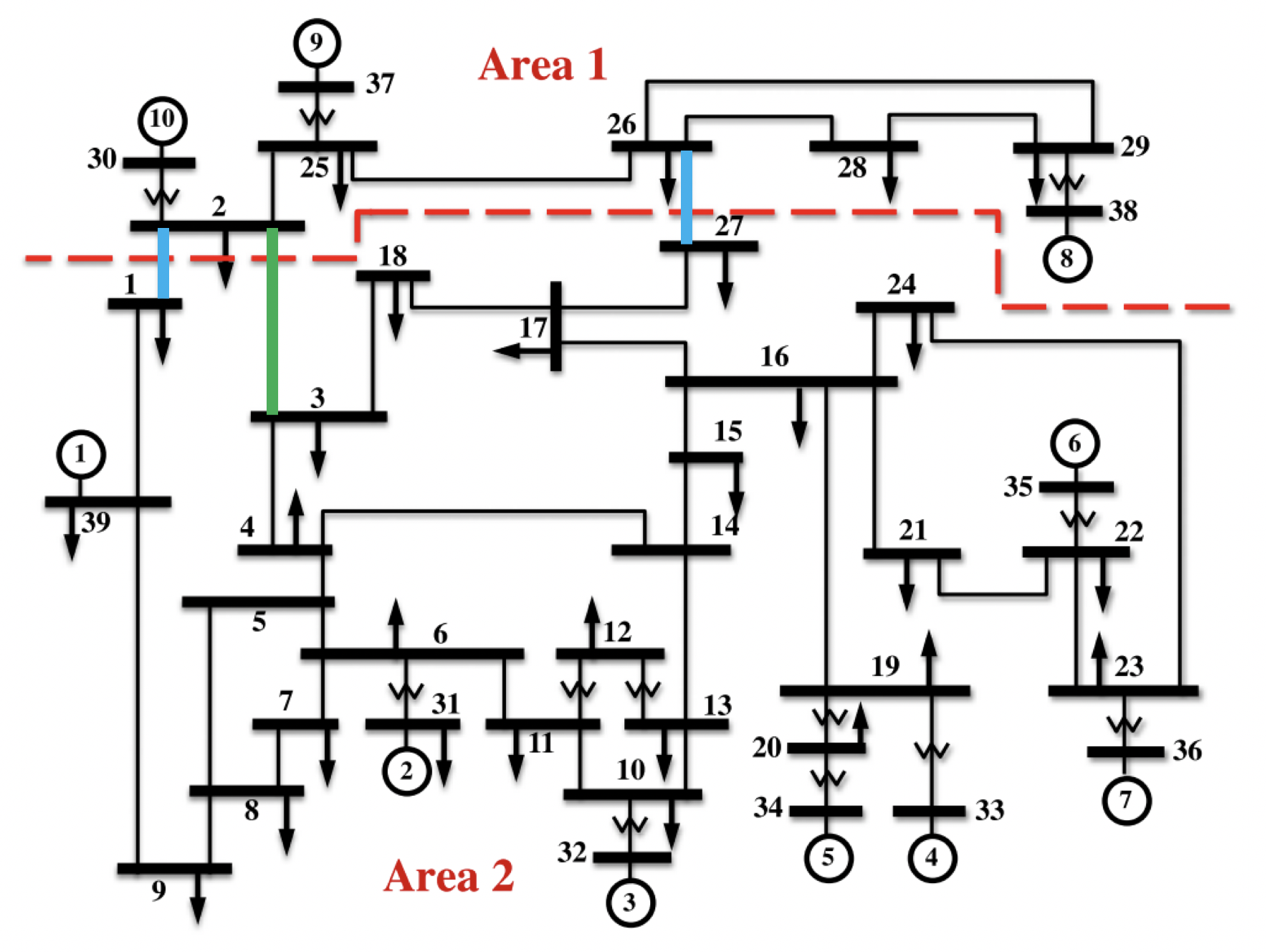}
\caption{IEEE 39-bus network with two control areas from \cite{zhao2016unified}. The two blue tie lines are switched off to create a tree structure with $(2,3)$ as bridge.}\label{fig:grid_39}
\end{figure}

We switch off the tie lines $(1,2)$ and $(26,27)$, chosen heuristically as those with the smallest absolute line flow. Two control areas are then connected in a tree structure as shown in Figure~\ref{fig:grid_39}. 
We implement the unified controller on all nodes as follows. We only allow to adjust generations at first, but, if a severe failure is detected, the controller is then allowed to reduce loads. As a last resort, zero area control error can be lifted. The threshold is set to $0.5$pu for dual variables.

\begin{figure}[!h]
\centering
\subfloat[Dual variable dynamics\label{fig:non_severe_dual}] {\includegraphics[width=.24\textwidth]{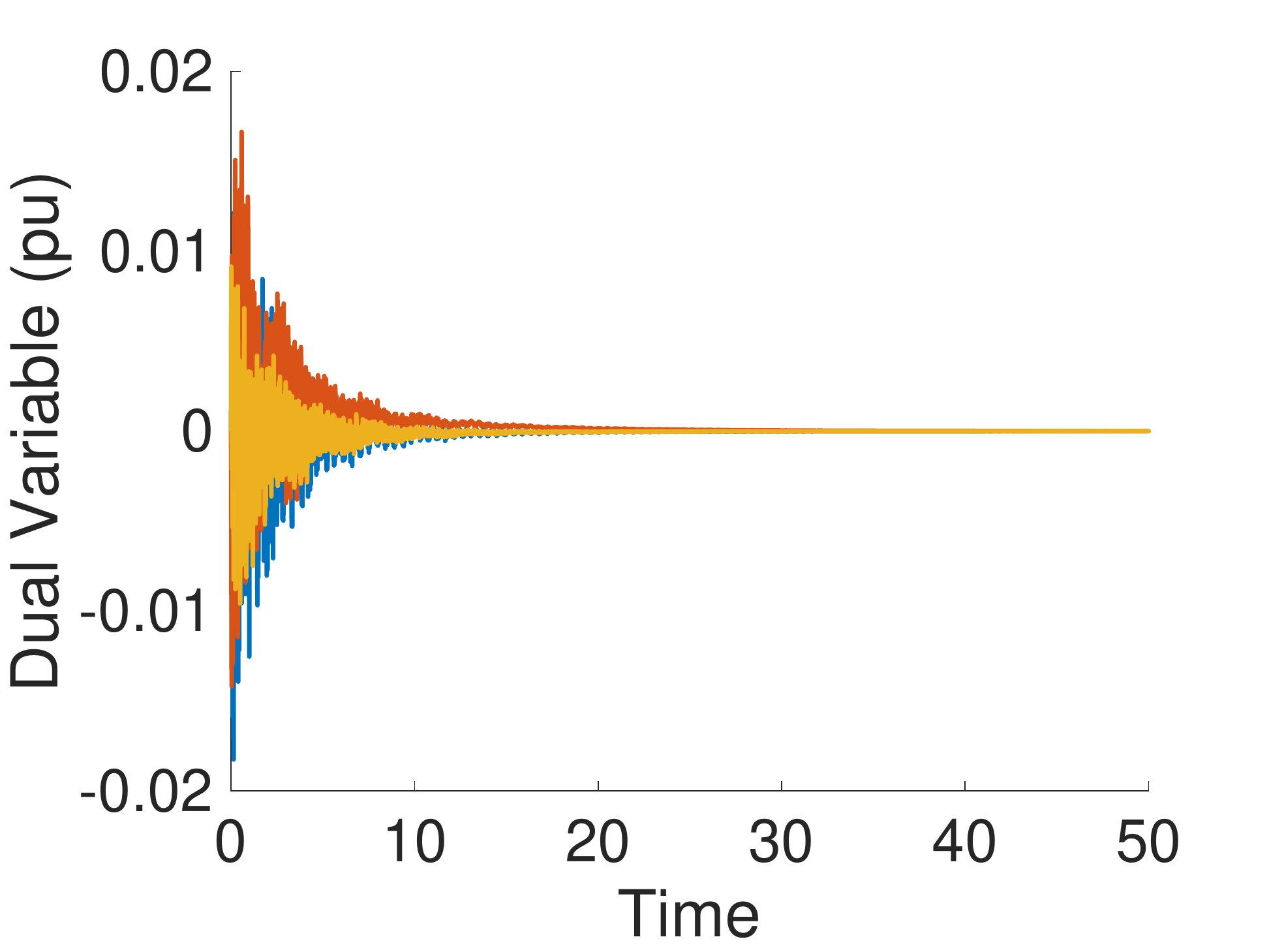}}
\hfill
\subfloat[Line flow dynamics\label{fig:non_severe_flow}] {\includegraphics[width=.24\textwidth]{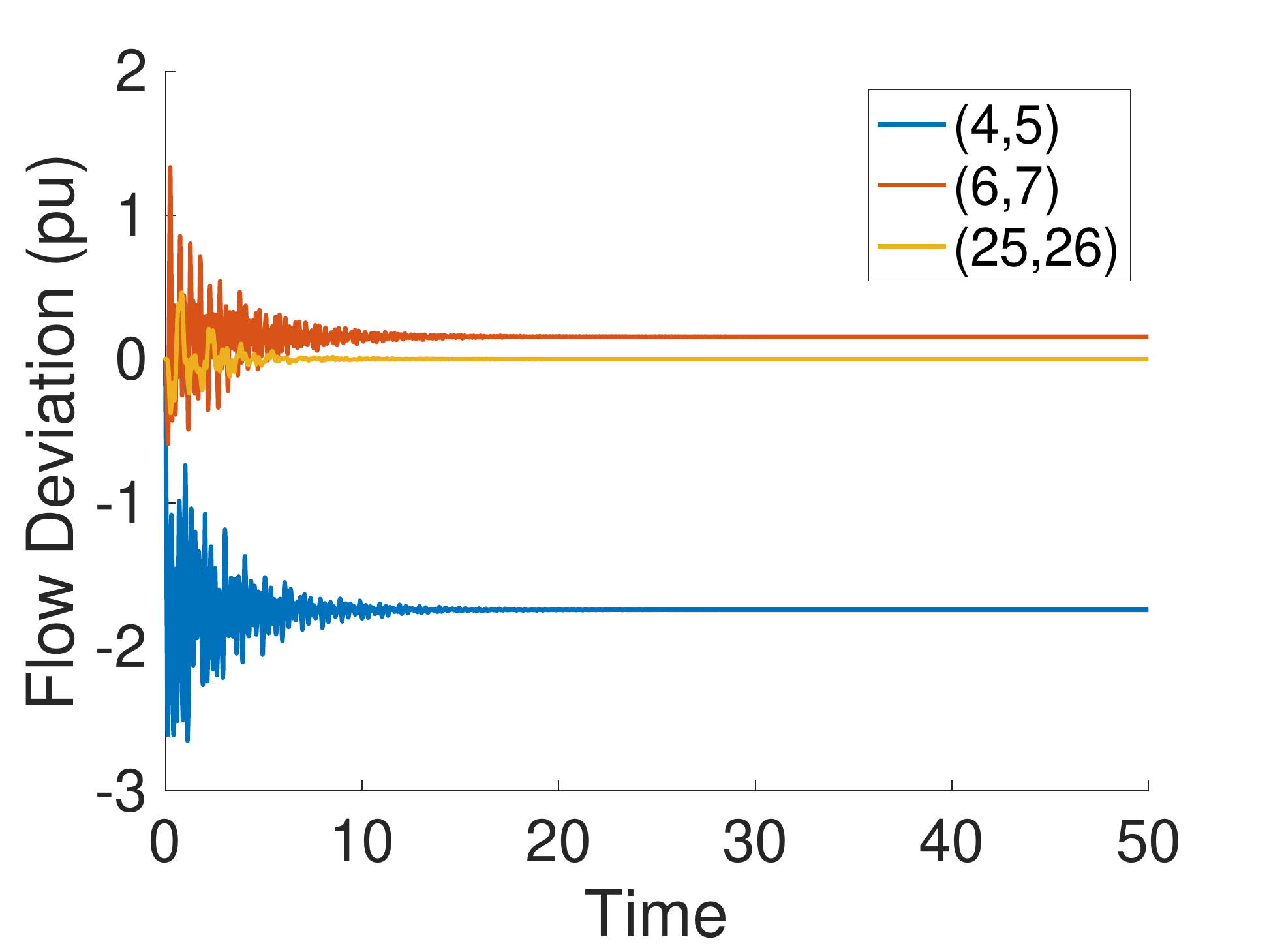}}
\caption{System dynamics after the non-severe failure of line $(4,14)$. The controller is only allowed to reduce generations.}
\label{fig:non_severe}
\end{figure}

\begin{figure}[!h]
\centering
\subfloat[Dual variable dynamics\label{fig:severe_dual}] {\includegraphics[width=.24\textwidth]{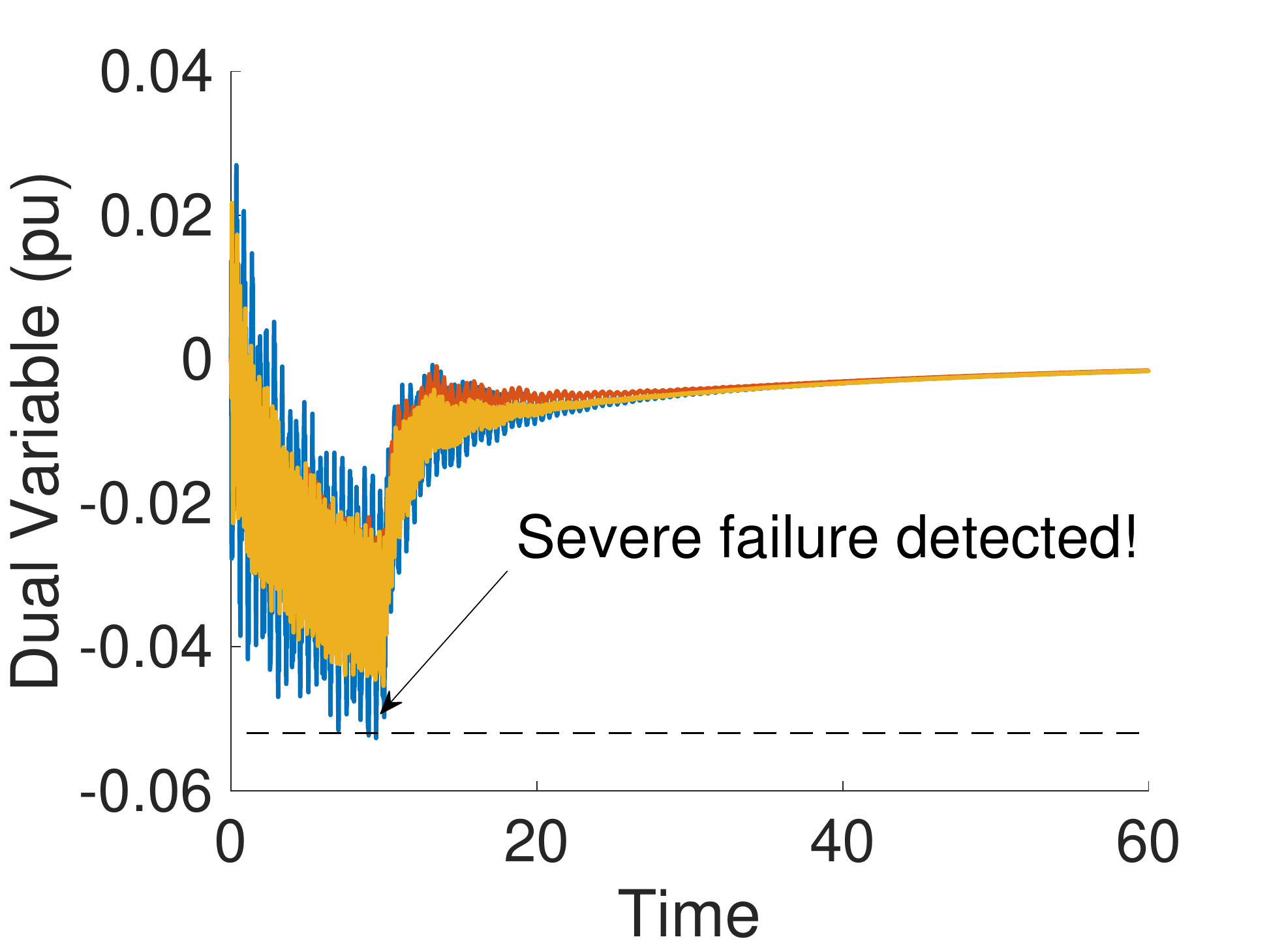}}
\hfill
\subfloat[Line flow dynamics\label{fig:severe_flow}] {\includegraphics[width=.24\textwidth]{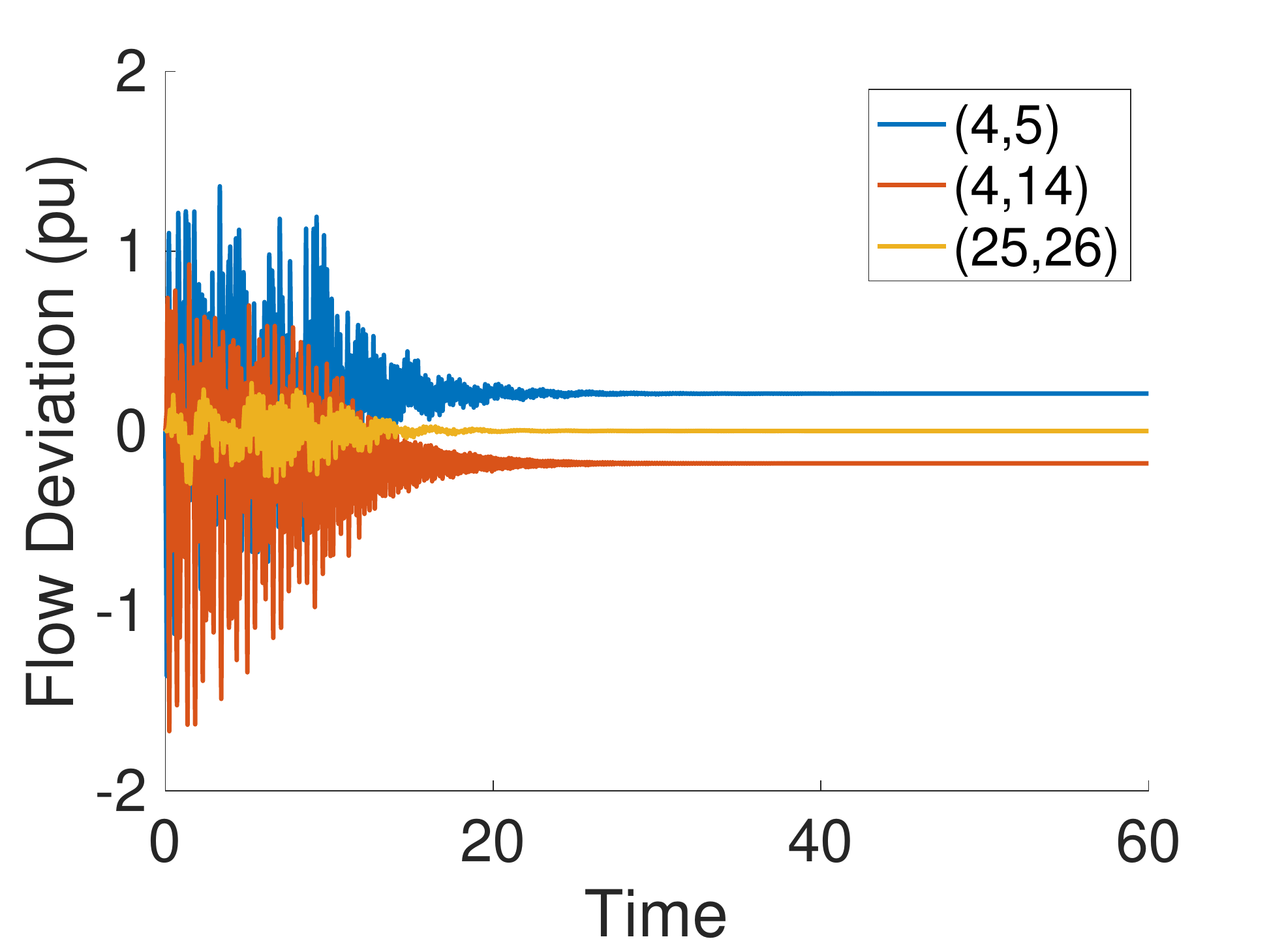}}
\caption{System dynamics after the severe failure of line $(6,7)$. A warning is raised at time $10$ sec, at which point the controller is allowed to shed loads.}
\label{fig:severe}
\end{figure}

As illustrated in Figure~\ref{fig:non_severe}, in the case of the non-severe failure $(4,14)$, the dual variables are always below the threshold and the system quickly converges to a safe equilibrium. On the other hand, the failure $(6,7)$ leads to unstable oscillations of dual variables and a severe warning is raised at $10$ sec, as depicted in Figure~\ref{fig:severe}. The controller is then allowed to shed loads, action that quickly re-stabilizes the system. Note that the flow on line $(25,26)$ remains unchanged at steady state for both failures, as it belongs to a non-associated control area.


\section{Simulations}\label{section:case_studies}
\subsection{Simulation Setup}
We evaluate our integrated approach on the IEEE 118-bus, 179-bus, 200-bus and 240-bus networks. For each network, we first partition it into two control areas using network modularity clustering algorithm as proposed in \cite{zocca2019or}. The tie line with largest absolute flow is selected as the remaining bridge, while all other tie lines are switched off. More information on these test networks are provided in Table~\ref{table:stat_network}.
\begin{table}[!t]
\def\arraystretch{1.3}
\centering
\caption{Statistics on IEEE test networks.} \label{table:stat_network}
\begin{tabular}{|c|c|c|c|}
\hline
Network & \# of Edges & Control Area Sizes & \# of Tie Lines  \\
\hline
118 & 186 & (83, 35) & 4 \\
\hline
179 & 263 & (98, 81) & 4\\
\hline
200 & 242 & (103, 94) & 9 \\
\hline
240 & 448 & (143, 97) & 10\\
\hline
\end{tabular}
\end{table}

The simulated scenarios are created as follows. For each network, we first adopt the nominal load profile from \cite{babaeinejadsarookolaee2019power} and obtain the generation profile through a DC OPF on the original network. The power injections from this DC OPF problem are used for both the original network and the network after line switching, while the resulting nominal line flows will generally be different. Next, we iterate over every transmission line as the initial single-line failure (tie lines that have been switched off are ignored)
and simulate the cascading process under four control approaches:
\begin{enumerate}
    \item Unified controller on tree-connected areas;
    \item Unified controller on mesh-connected areas;
    \item Automatic generation control (AGC)\footnote{We model AGC as an algorithm that solves \eqref{eqn:uc_olc} without flow limits \eqref{eqn:line_limit}.} on tree-connected areas;
    \item Automatic generation control on mesh-connected areas.
\end{enumerate}
For presentation simplicity, we refer these approaches as UC+T, UC+M, AGC+T and AGC+M.
Moreover, to test their performance under various network congestion conditions, we scale the line flow limit and generation limit by a common factor $\alpha=0.5, 1, 1.5$. For a fair comparison, all strategies follow the same relaxation procedure as described in the previous section when the optimization is infeasible. We remark that since AGC does not enforce line  limits, the cascading failure may proceed for multiple stages as described in Section~\ref{section:pre}, while UC always terminates the failure at the first stage.

We quantify the mitigation performance using the \textit{load loss rate} (LLR), defined as the ratio of the total loss of load to the total demand. The localization performance is evaluated by the \textit{adjusted generator rate} (AGR) which is the ratio of the number of generators whose operating points have changed after the failures to the total number of generators. The range for both LLR and AGR is between $[0,100\%]$, and a smaller value indicates better mitigation/localization performance. The complete results are summarized in Tables~\ref{table:stat1} and \ref{table:stat2} for reference. In the following subsections, we highlight several insights from these results.

\subsection{Performance of Our Approach}
We first evaluate the performance of our integrated approach UC+T, and compare it with the traditional approach AGC+M. Table~\ref{table:uc_t_agc_m} shows the fraction of failure scenarios with load loss (LL) and adjusted generations (AG), as well as the average LLR and AGR of those scenarios for these two control approaches. Each cell contains three values, representing the results for $\alpha=0.5, 1, 1.5$ respectively.

In terms of failure mitigation, our proposal prevents load shedding from happening in more failure scenarios over 118-bus, 179-bus, and 240-bus networks under almost all network congestion conditions. More importantly, our approach always achieves higher grid reliability, in the sense that the average load loss, for failures where load shedding is inevitable, is uniformly lower compared with the traditional AGC+M approach. This improvement is particularly significant when the grid becomes congested (smaller $\alpha$). 
The average LLR, averaged over all failure scenarios with load loss, is 3\% or lower under UC+T but can be as high as 19\% under AGC+M.
{Moreover, large-scale outages with load loss larger than $10\%$  almost never happen under our proposed UC+T approach.}


In terms of failure localization, we observe that on 118-bus, 179-bus and 240-bus networks,  generators under UC+T are adjusted for more failure scenarios when the network is less congested, while achieving less load loss compared to the traditional approach. When the congestion becomes worse, the traditional AGC+M approach not only leads to more scenarios with adjusted generators, but results in more load loss as well. This fact indicates that generators under UC+T approach adjust their operations more actively in response to failures. Moreover, generators react more efficiently and locally under UC+T, as the average AGR is always lower.

It should be noted that there are more failure scenarios with load loss or generator adjustment for our approach over the 200-bus network. Nevertheless, the LLR and AGR are still uniformly lower on average under UC+T. One possible explanation might be that the 200-bus network is very sparse (see Table~\ref{table:stat_network}), with limited power transfer capacity across control areas and, hence, more prone to potential failures. By adjusting the generators and loads more proactively, UC+T tries to make sure the post-contingency injections are more robust in the new topology. This, again, confirms that UC+T tends to adjust generators more actively to failures, and the adjustment is prescribed in a more efficient and local manner.

\begin{table}[!t]
    \def\arraystretch{1.2}
    \centering
    \caption{Statistics on UC+T (first row) and AGC+M (second row)\protect\footnotemark.} \label{table:uc_t_agc_m}
    \begin{tabular}{|c|c|c|c|c|}
        \hline
         Network &  118 & 179 & 200 & 240 \\
         \hline
        \multirow{2}{*}{\shortstack[c]{\% of Scen. \\ w/ LL}} & $\mathbf{49, 8}, 5$ & $\mathbf{4, 2}, 2$ & $\mathbf{21}, 19, 15$ & $\mathbf{22, 9, 3}$ \\ 
        \cline{2-5}
         & $99, 26, \mathbf{4}$ & $59, 4, \mathbf{1}$ & $24, \mathbf{8, 7}$ & $98, 95, \mathbf{3}$\\
         \hline
         \multirow{2}{*}{\shortstack[c]{\% of Scen. \\ w/ AG}} & $\mathbf{94}, 53, 8$ & $\mathbf{71, 44}, 37$ & $56, 47, 31$ & $\mathbf{95, 74}, 22$\\ 
        \cline{2-5}
         & $99, \mathbf{27, 7}$ & $84, 63, \mathbf{20}$ & $\mathbf{37, 20, 21}$ & $98, 95, \mathbf{16}$\\
         \hline
        \multirow{2}{*}{\shortstack[c]{Avg. \\ LLR (\%)}} & $\mathbf{2, 2, 1}$& $\mathbf{2, 2, 2}$ & $\mathbf{3}, 2, 1$ & $\mathbf{1, 1, 2}$\\
        \cline{2-5}
         &  $19, 5, 4 $& $5, 3, 2$ & $14, \mathbf{2, 1}$ & $10, 6, 3$\\
         \hline
         \multirow{2}{*}{\shortstack[c]{Avg. \\ AGR (\%)}} & $\mathbf{10, 12, 16}$ & $\mathbf{27, 27, 37}$ & $\mathbf{30, 36, 51}$ & $\mathbf{24, 22, 52}$\\ 
        \cline{2-5}
         & $21, 15, 18$ & $74, 75, 39$ & $50, 36, 51$ & $60, 56, 63$\\
         \hline
    \end{tabular}
\end{table}
\footnotetext{We highlight in bold font the approach with better performance (two values may look the same after rounding). See Tables~\ref{table:stat1} and \ref{table:stat2} for more precise results.}

\subsection{Effects of the Unified Controller}
\begin{figure*}
\centering
\subfloat[118-bus network] {\includegraphics[width=.24\textwidth]{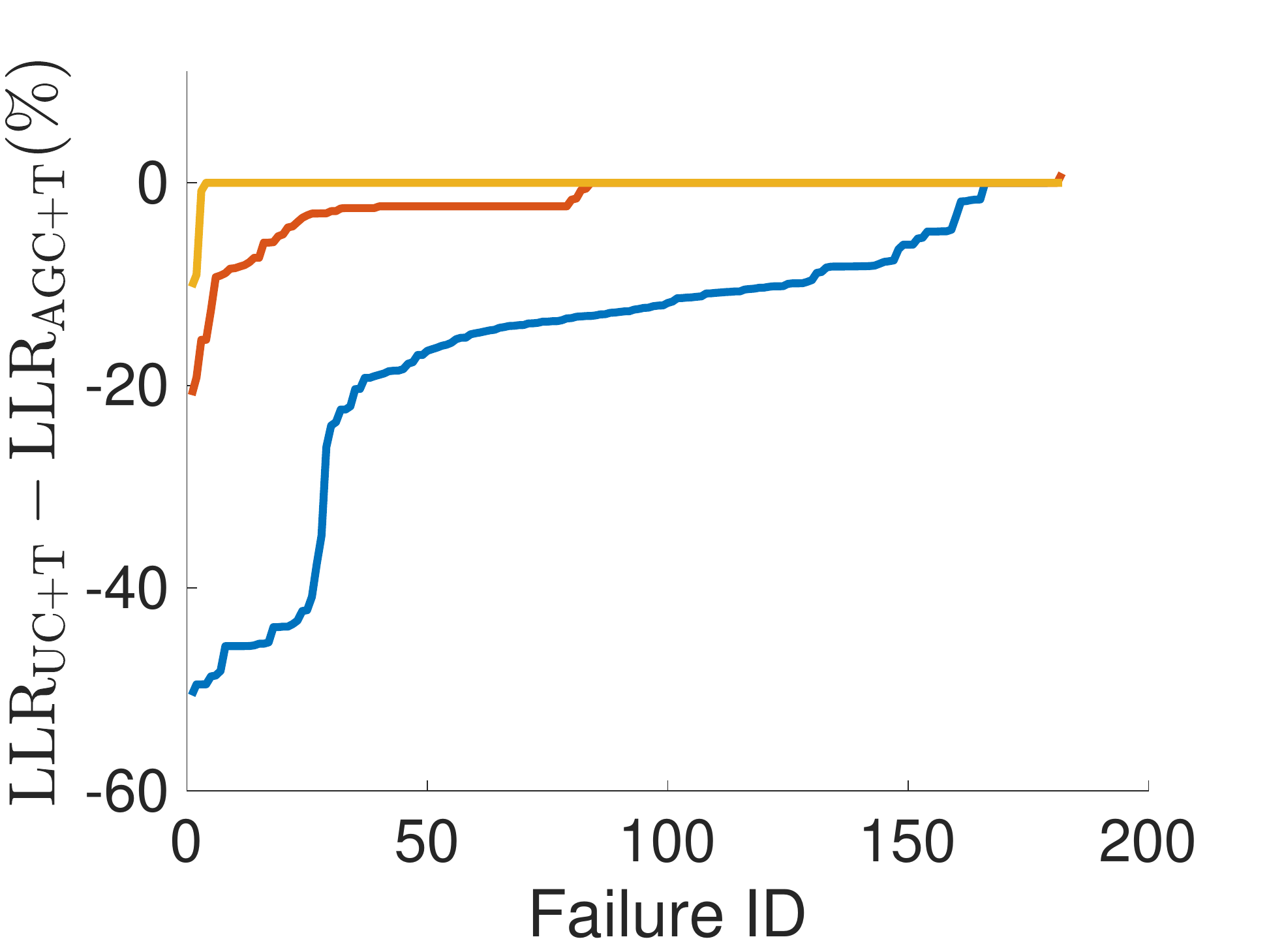}}
\subfloat[179-bus network] {\includegraphics[width=.24\textwidth]{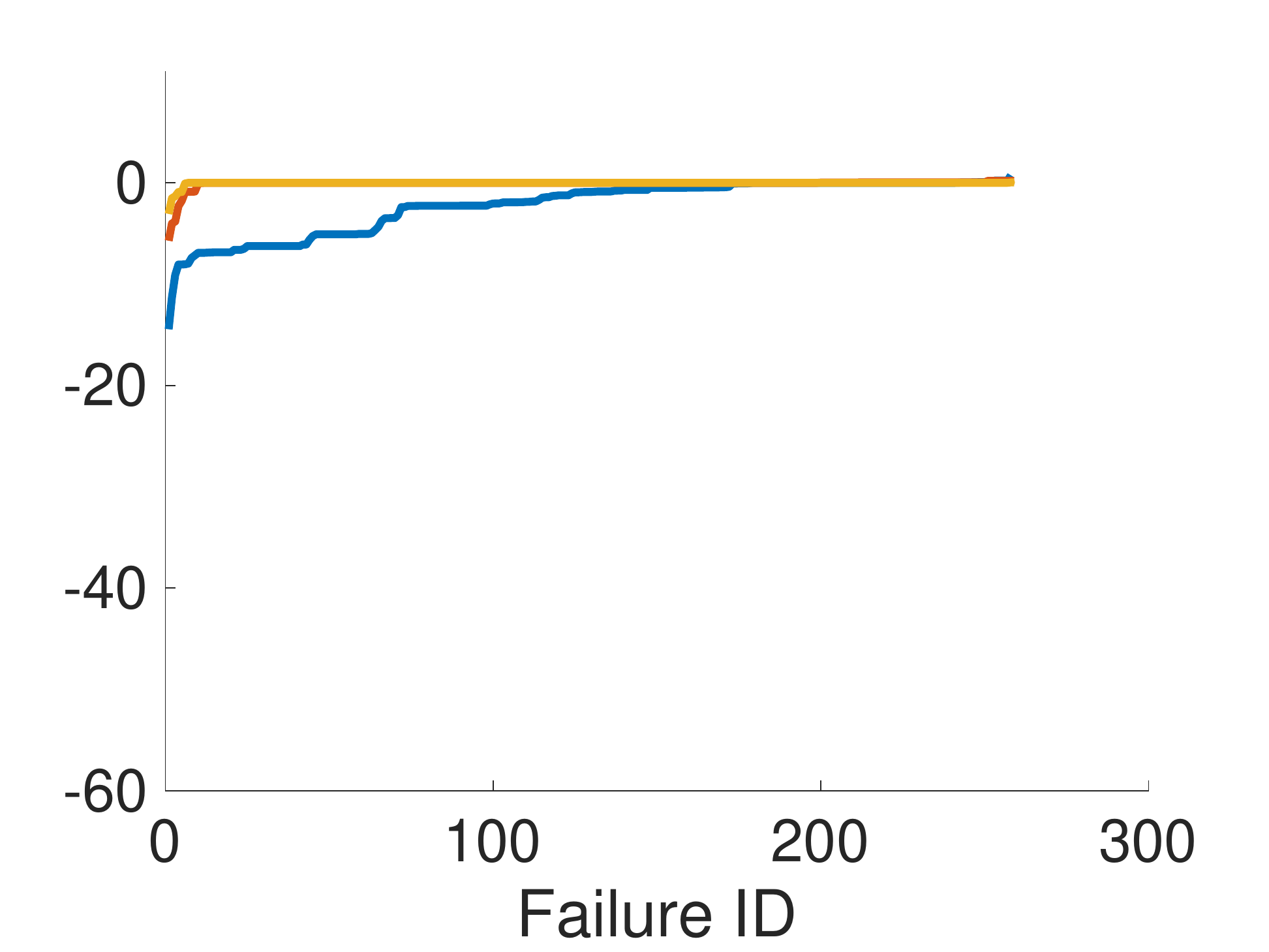}}
\subfloat[200-bus network] {\includegraphics[width=.24\textwidth]{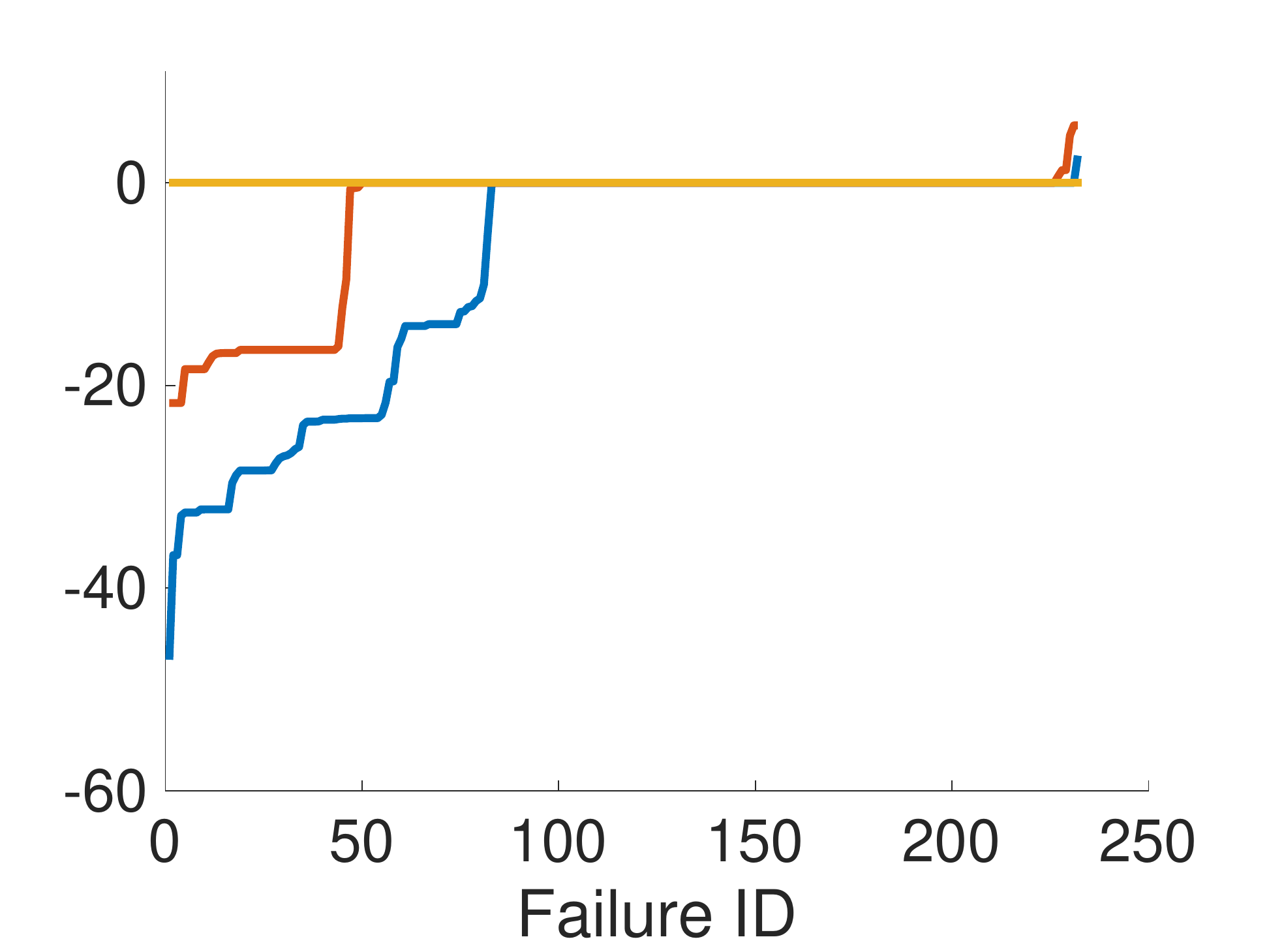}}
\subfloat[240-bus network] {\includegraphics[width=.24\textwidth]{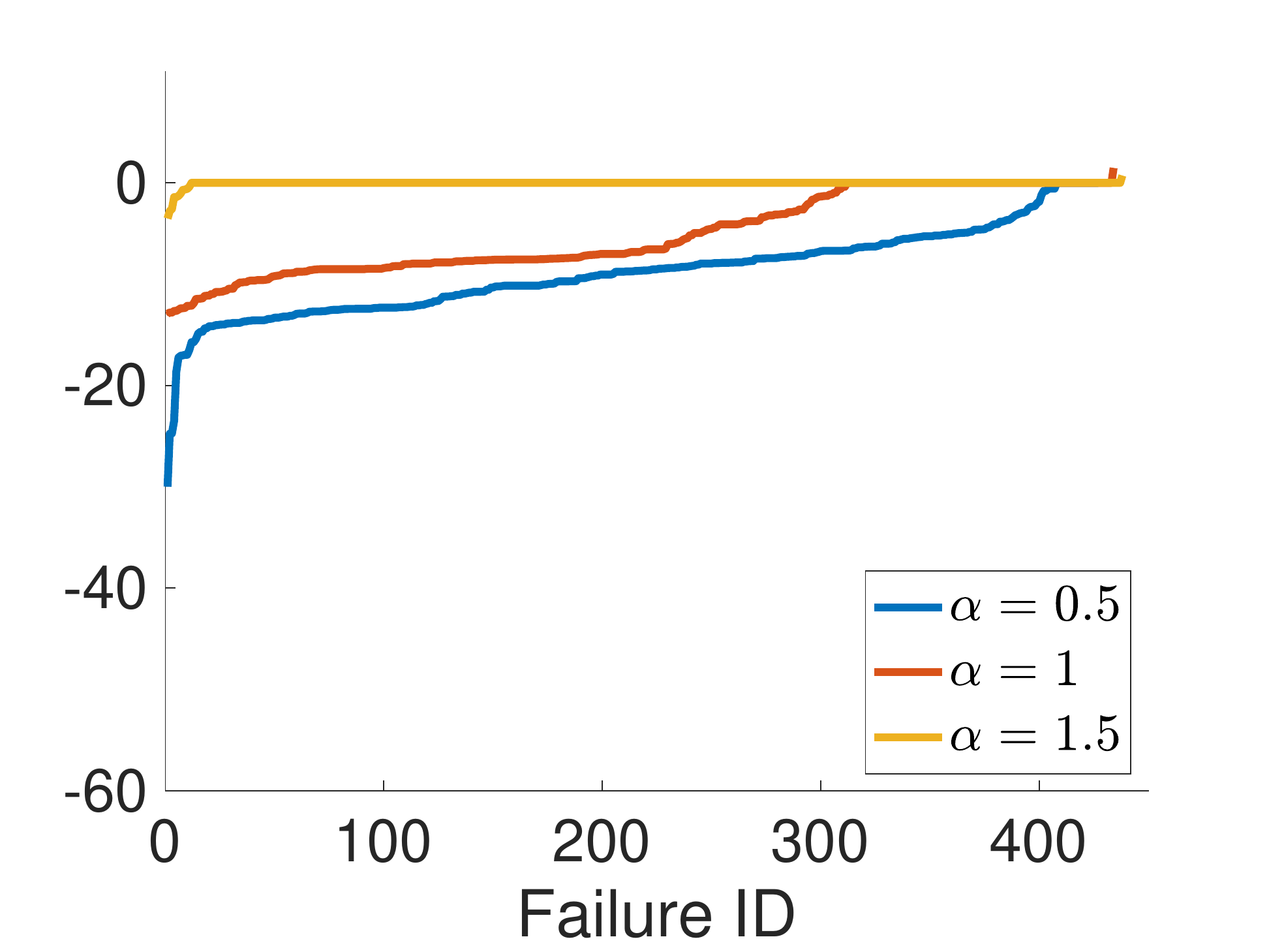}}\\
\subfloat[118-bus network] {\includegraphics[width=.24\textwidth]{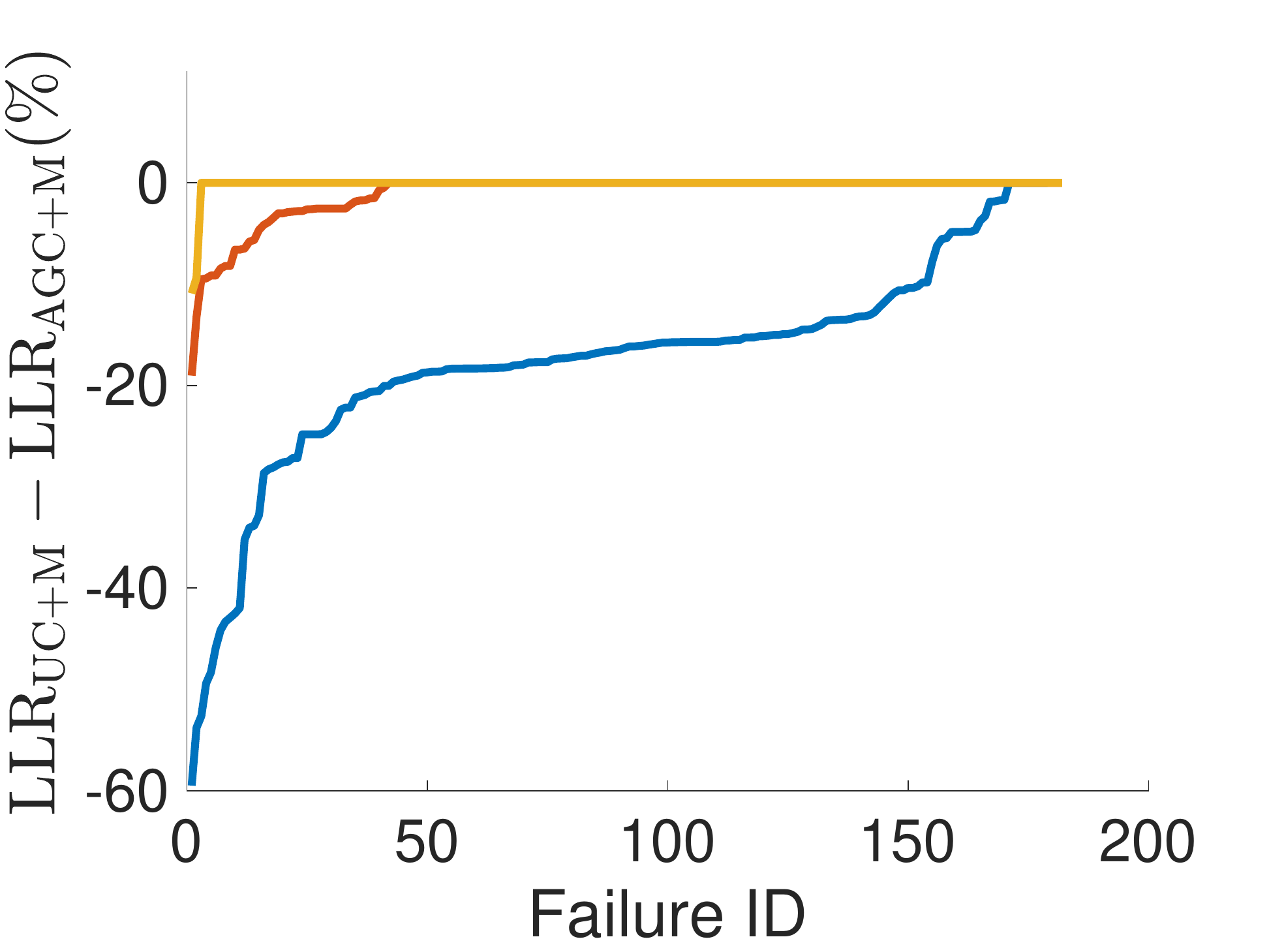}}
\subfloat[179-bus network] {\includegraphics[width=.24\textwidth]{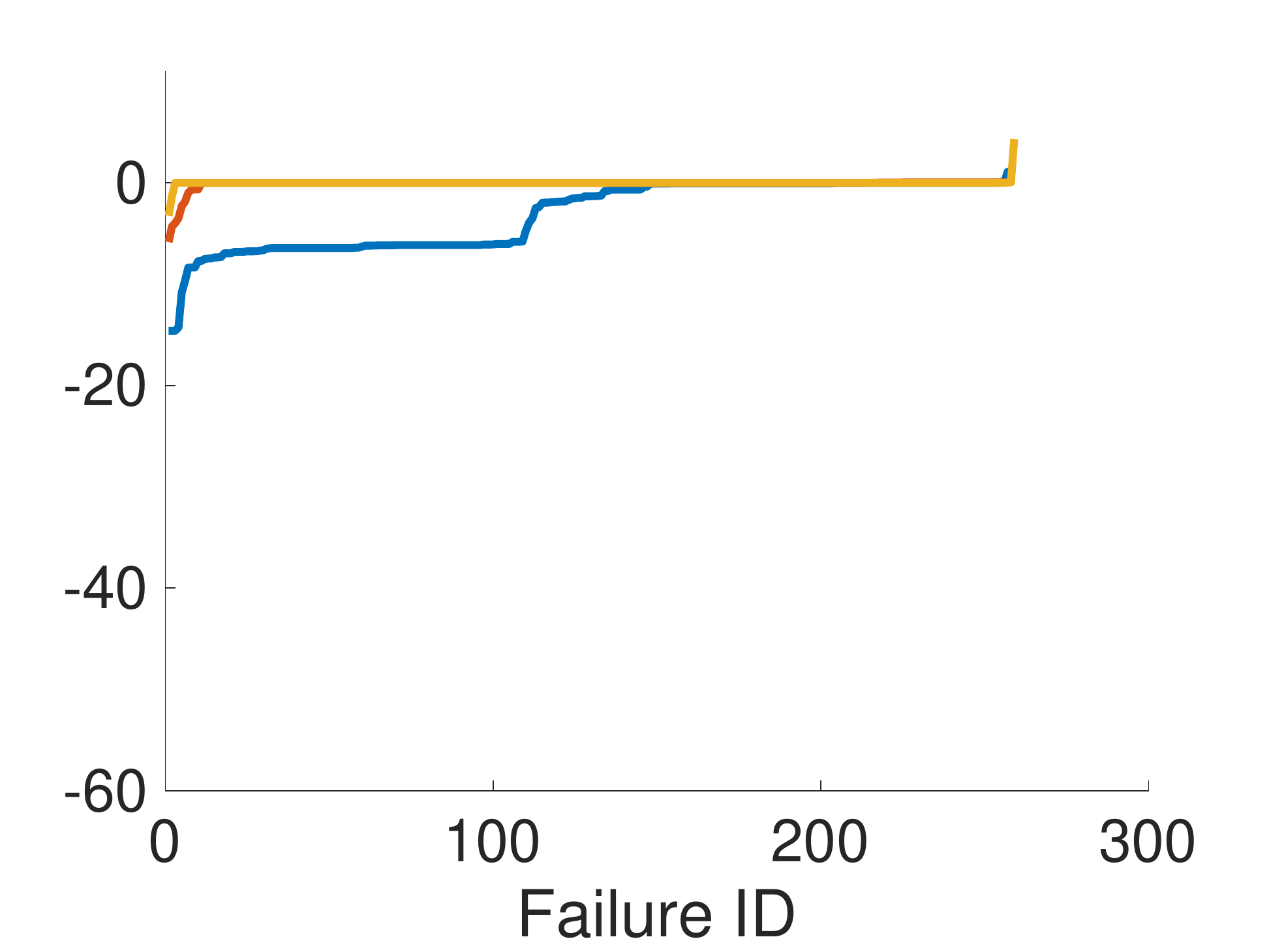}}
\subfloat[200-bus network] {\includegraphics[width=.24\textwidth]{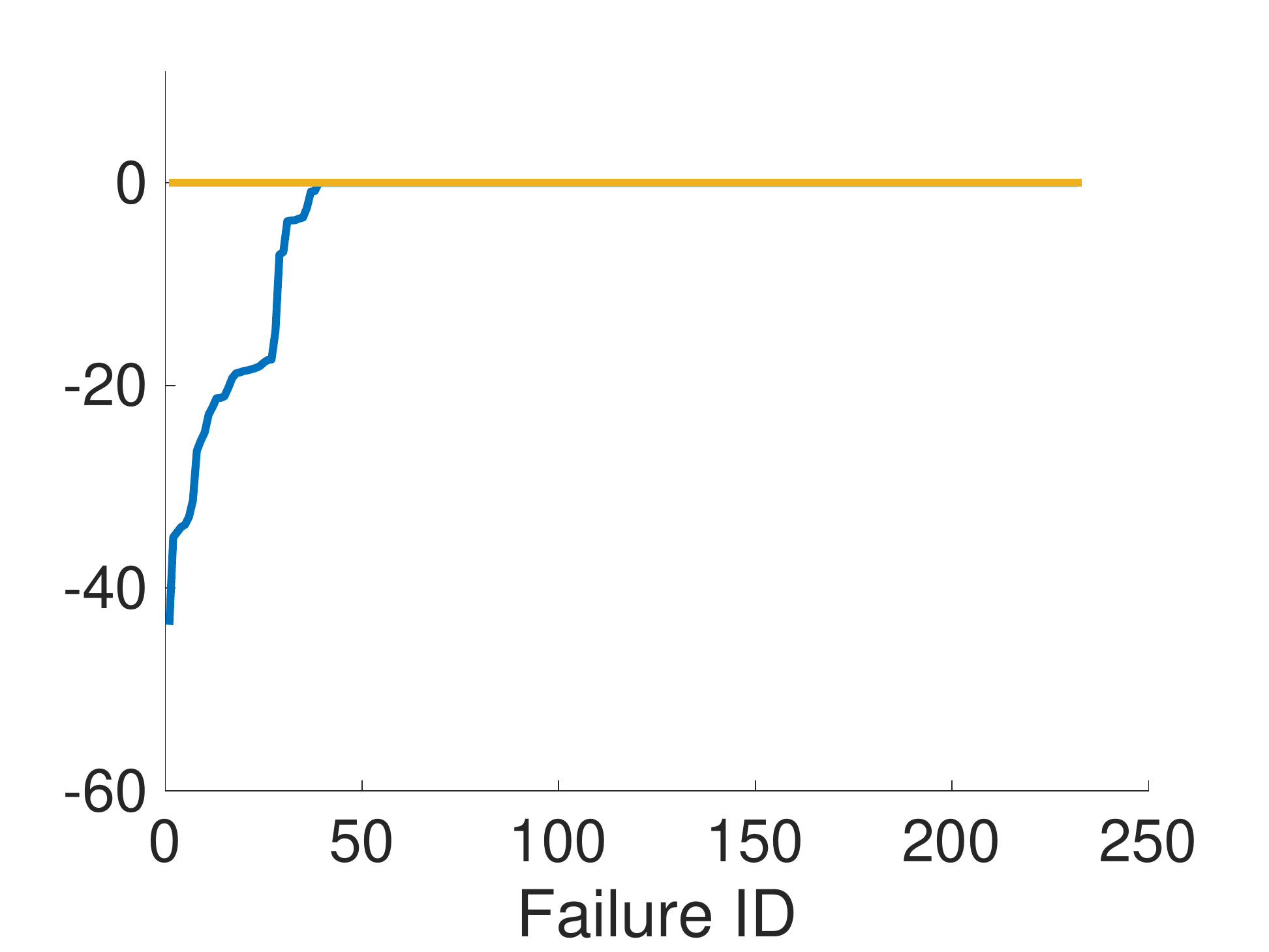}}
\subfloat[240-bus network] {\includegraphics[width=.24\textwidth]{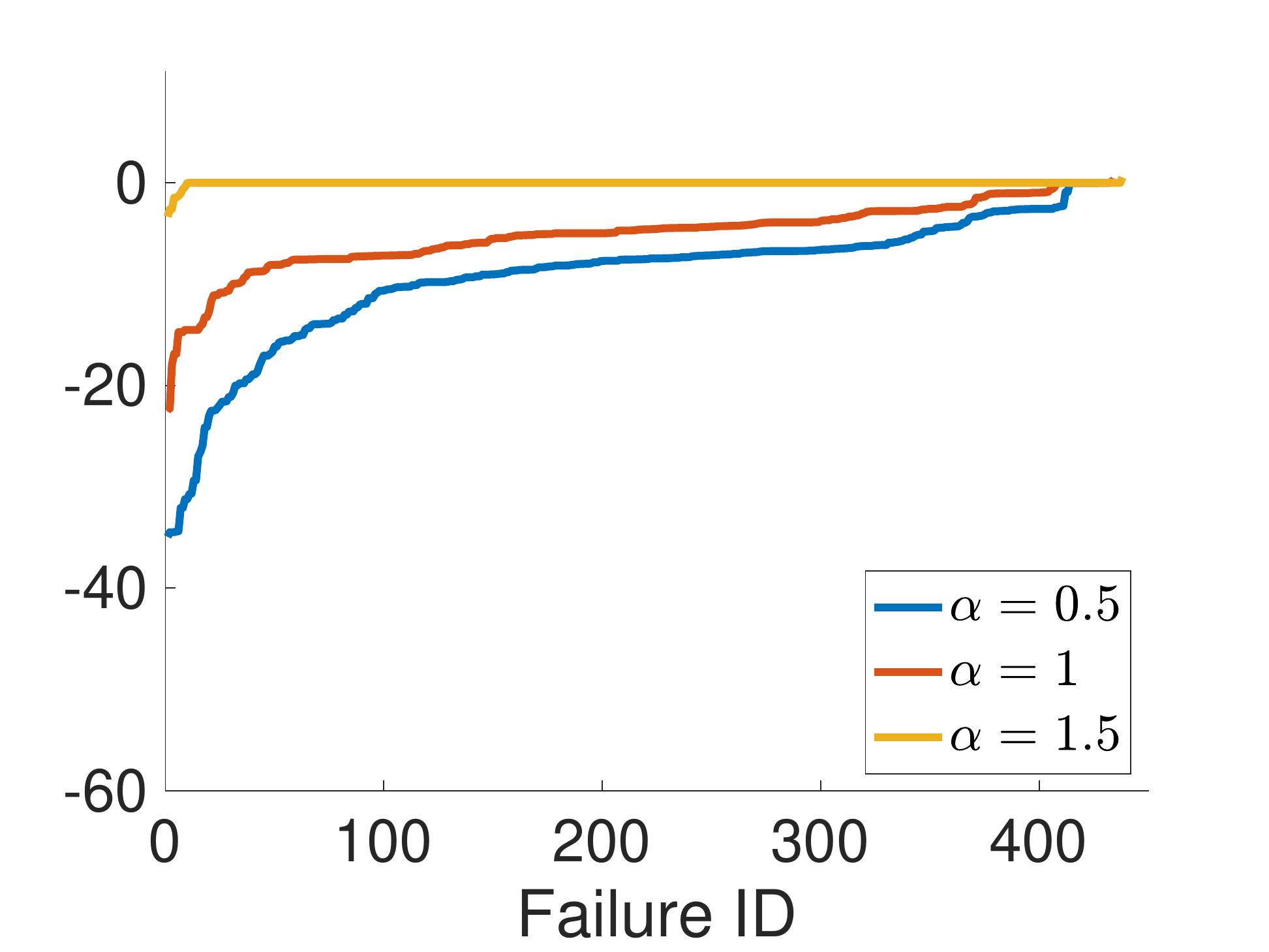}}
\caption{Difference of LLR between UC-based and AGC-based approaches.}
\label{fig:uc_agc}
\end{figure*}
The benefits of our approach stem from both UC that enforces line limits at fast timescale and tree-connected topology that localizes the impact of initial failures.  In this subsection and the next one we disentangle the benefit of each.

We first demonstrate the load loss reduction capability of UC by comparing UC-based and AGC-based approaches (UC+T vs. AGC+T and UC+M vs. AGC+M). For each pair of approaches, we compute the difference of LLR for every failure scenario and plot the results in Figure~\ref{fig:uc_agc}. Note that the failure index is re-ordered so that the difference of LLR is in a non-decreasing order for better presentation. 

The main insight from Figure~\ref{fig:uc_agc} is that UC-based approaches (whether the control areas are tree-connected or not) almost always reduce load loss, and such reductions are more pronounced as network congestion increases. 
We remark that there are indeed failure scenarios with less load loss under AGC-based approaches. When this happens, however, the failures always cascade through multiple stages under AGC and, hence, the time to re-stabilize the system is significantly longer. This suggests that UC+T prioritizes a shorter system stabilization time over load loss rate in the relatively rare cases where quickly stopping the cascading process leads to increased load loss.

\subsection{Effects of the Tree Partitioning}
\begin{figure*}
\centering
\subfloat[118-bus network] {\includegraphics[width=.24\textwidth]{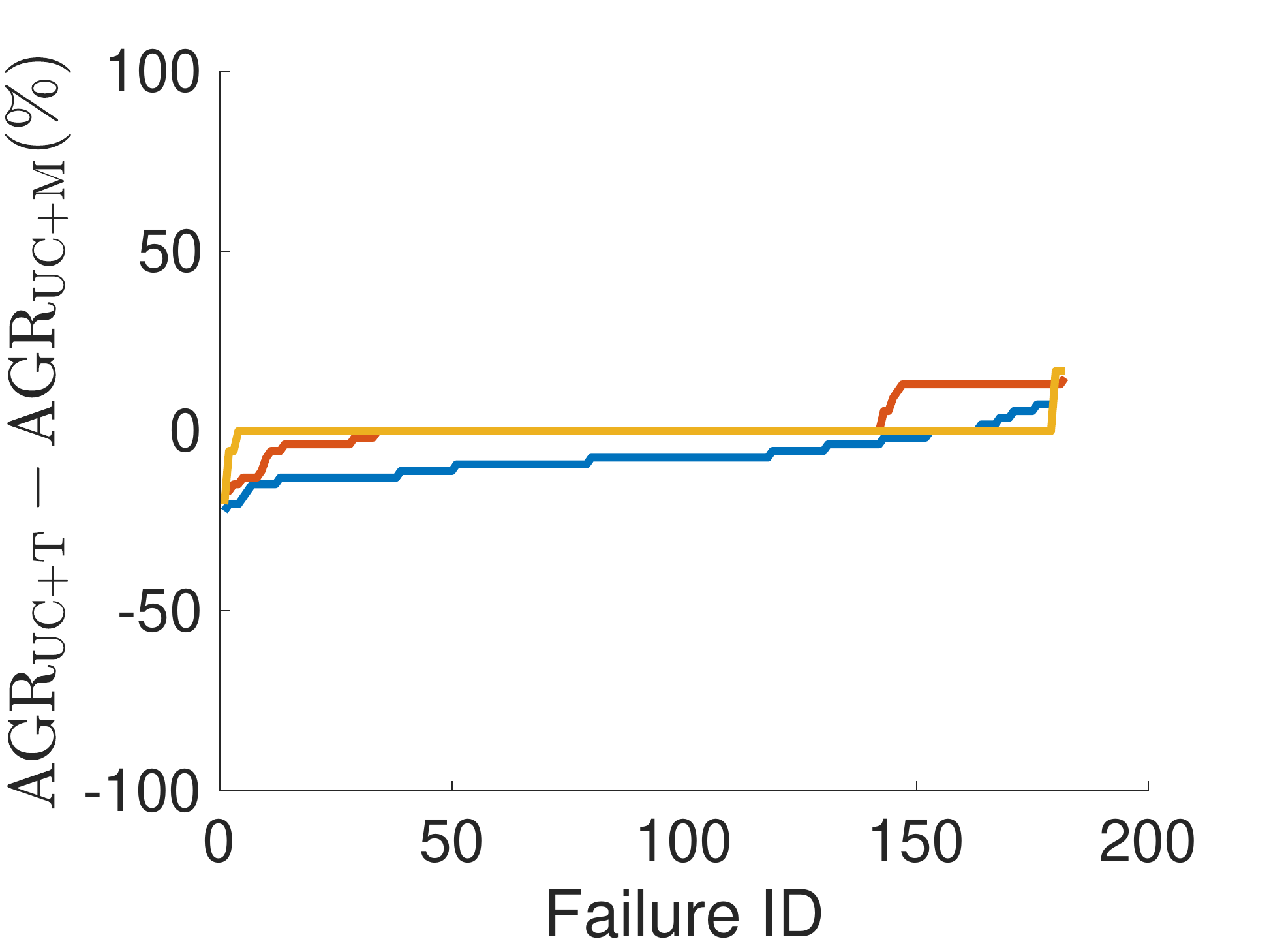}}
\subfloat[179-bus network] {\includegraphics[width=.24\textwidth]{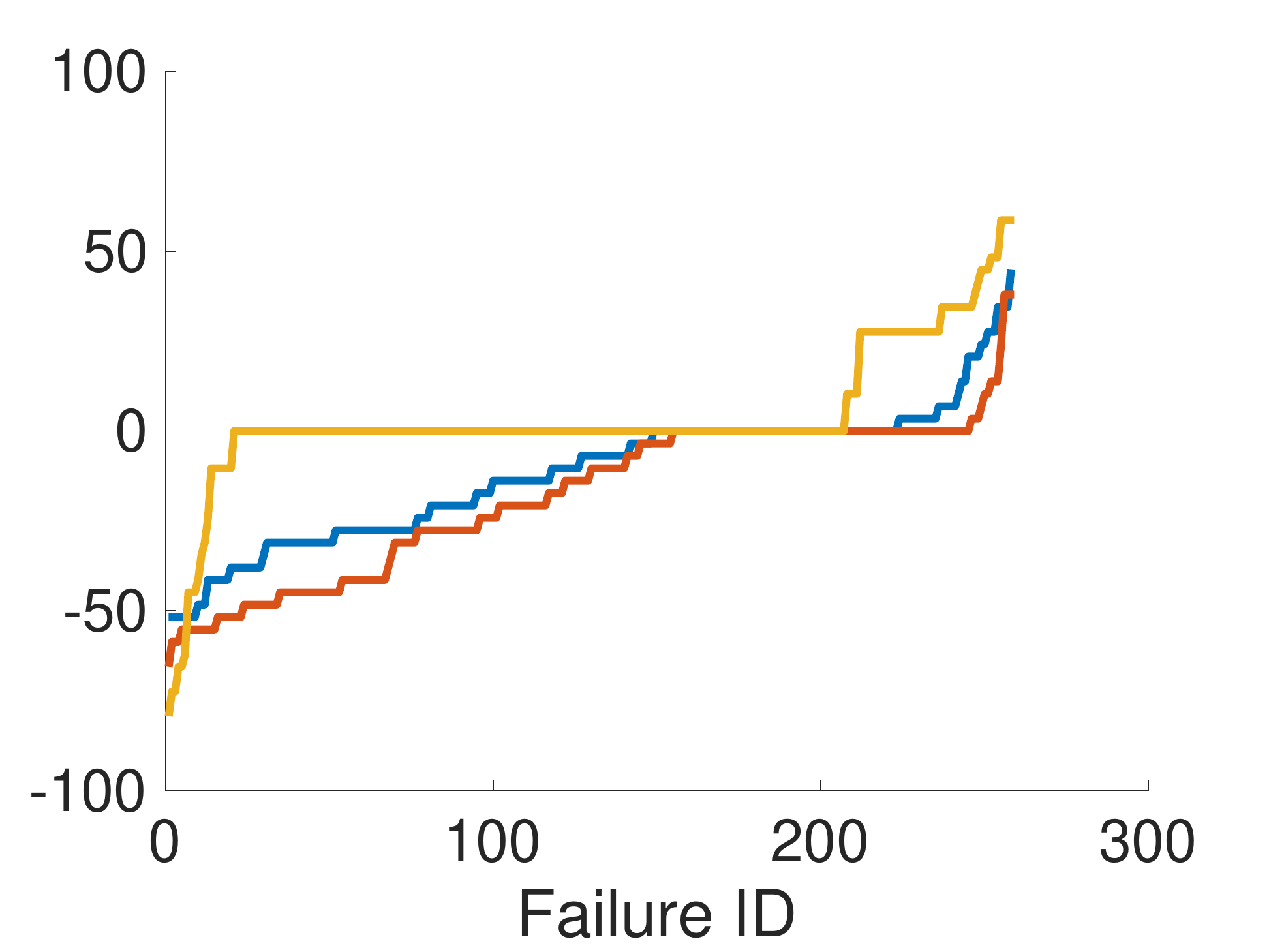}}
\subfloat[200-bus network] {\includegraphics[width=.24\textwidth]{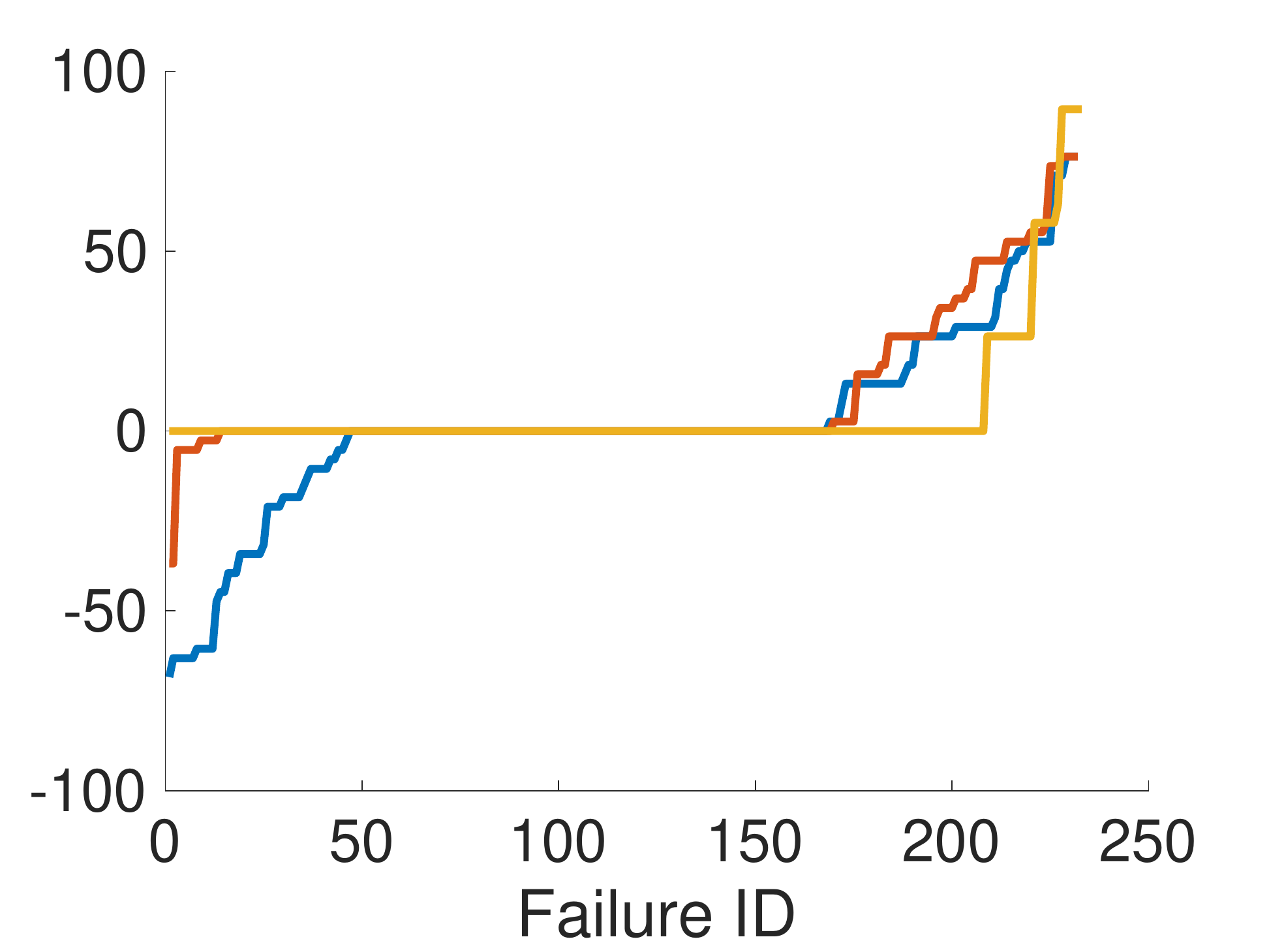}}
\subfloat[240-bus network] {\includegraphics[width=.24\textwidth]{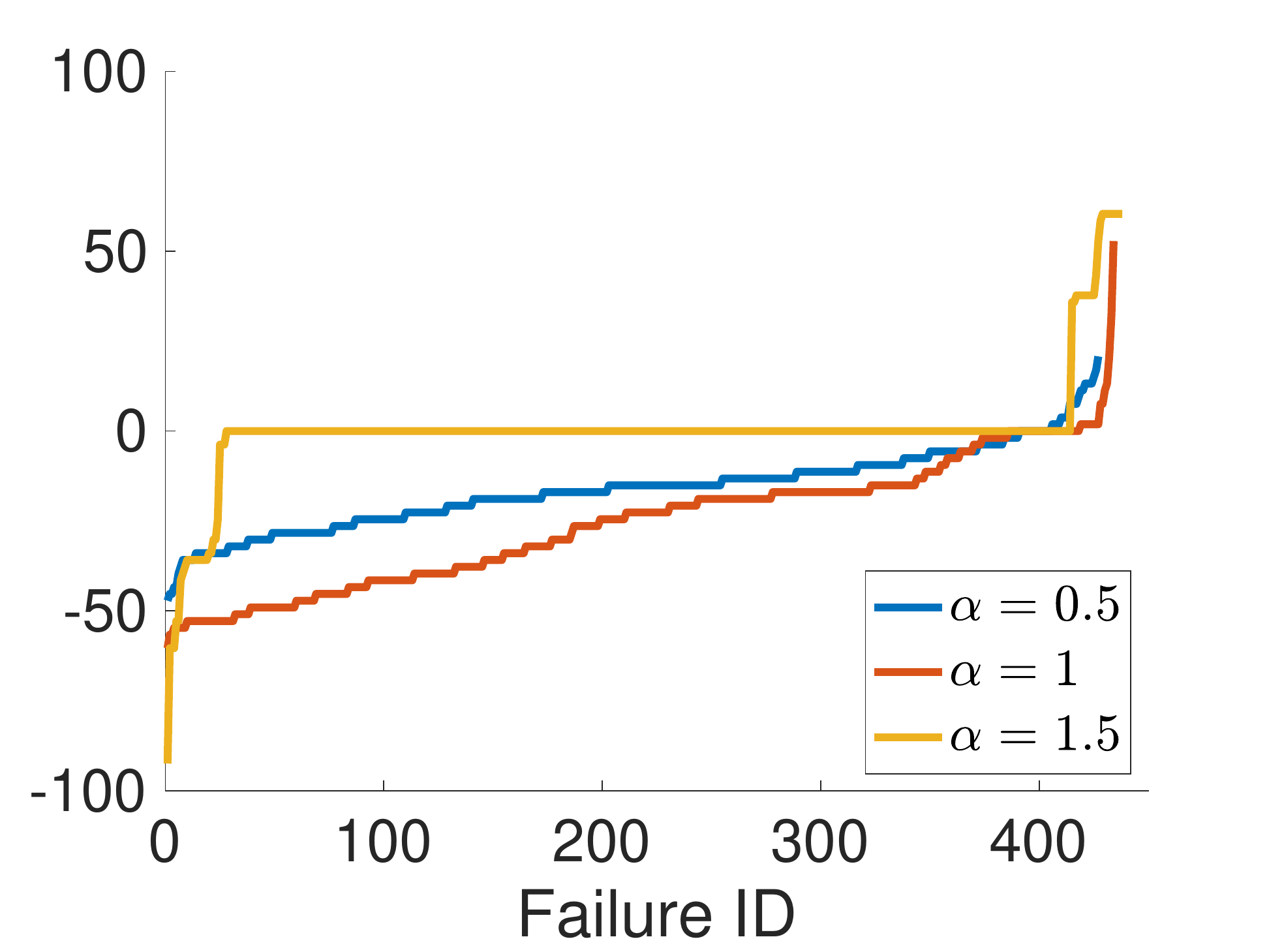}}\\
\subfloat[118-bus network] {\includegraphics[width=.24\textwidth]{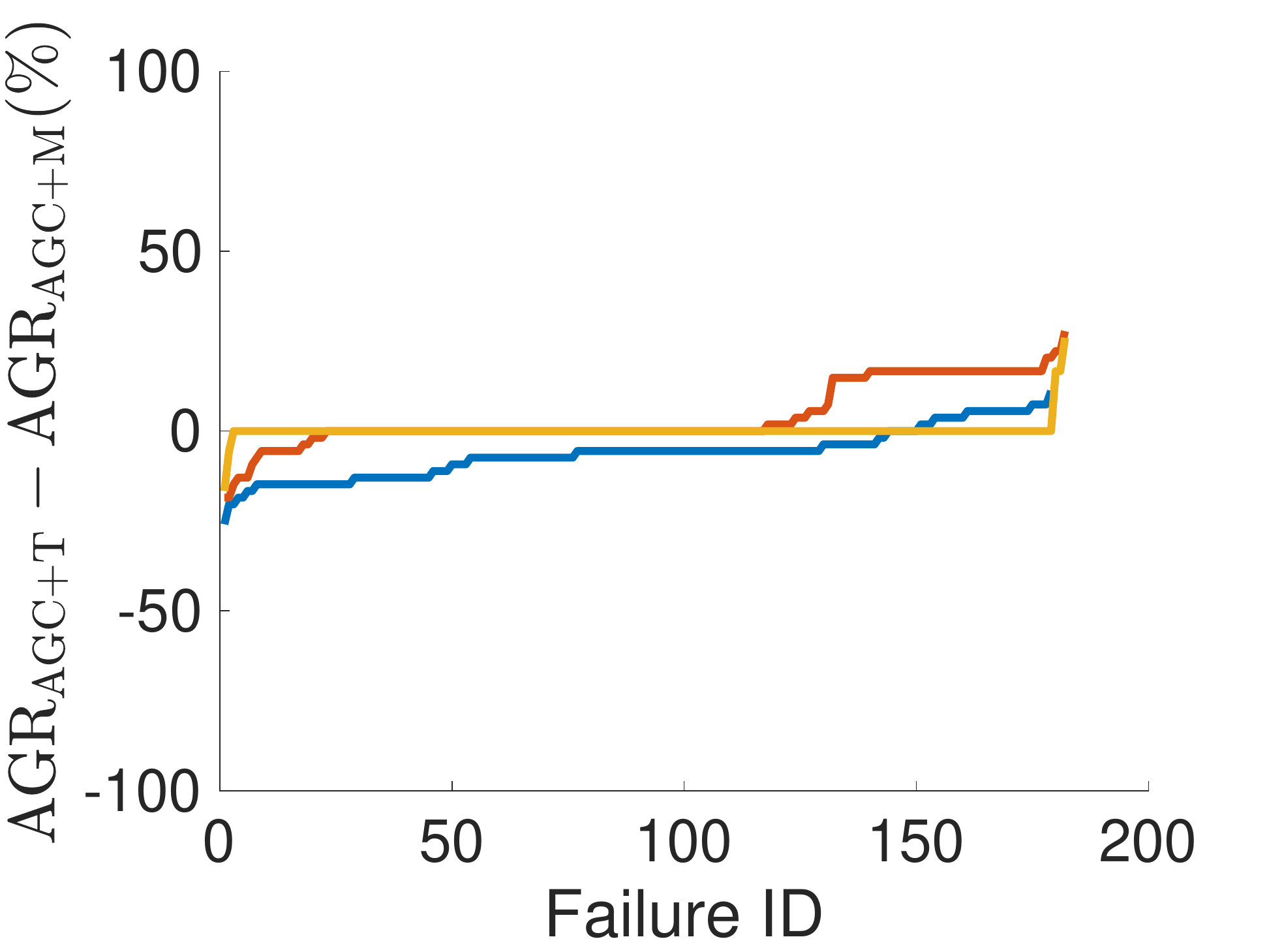}}
\subfloat[179-bus network] {\includegraphics[width=.24\textwidth]{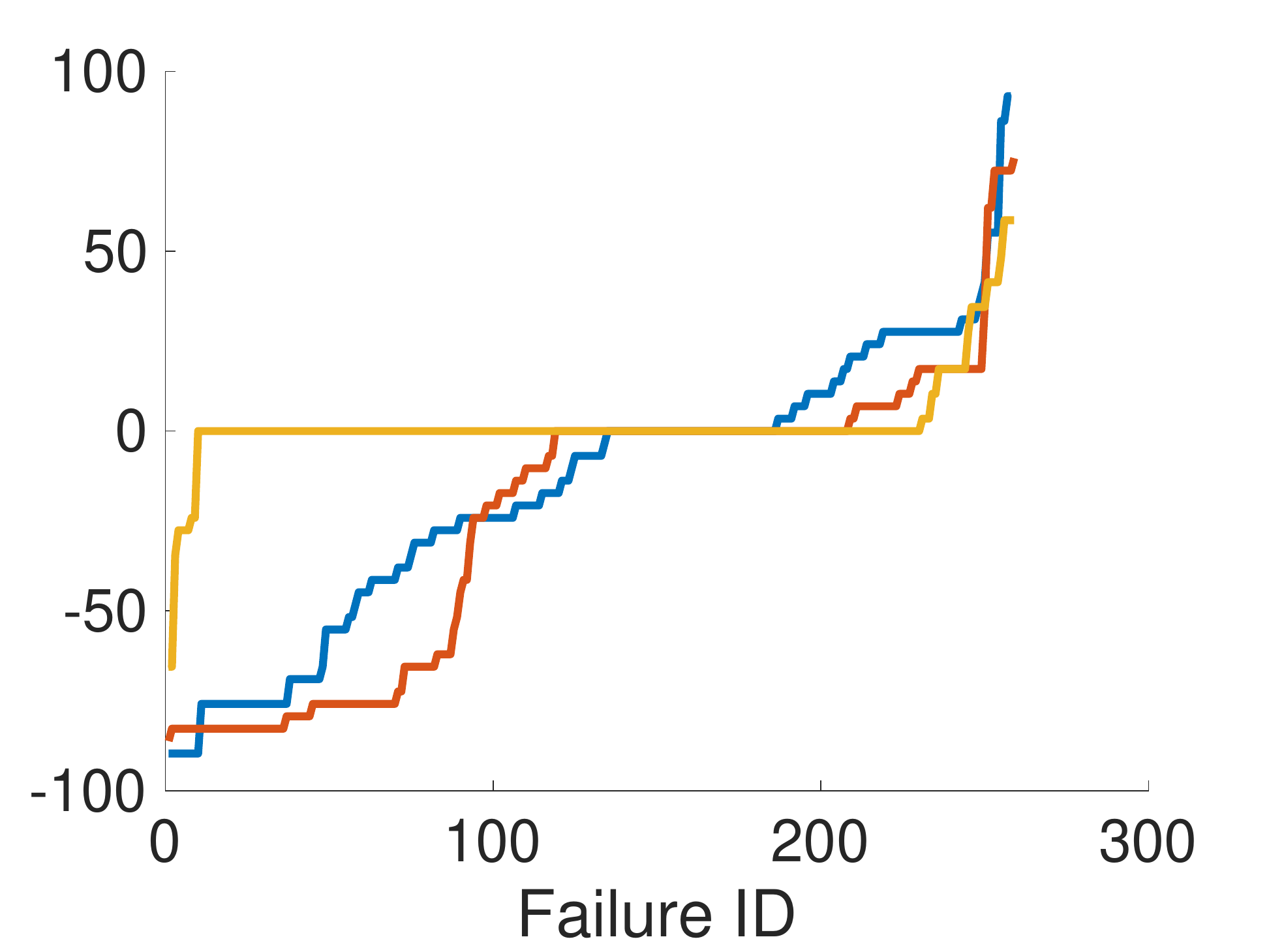}}
\subfloat[200-bus network] {\includegraphics[width=.24\textwidth]{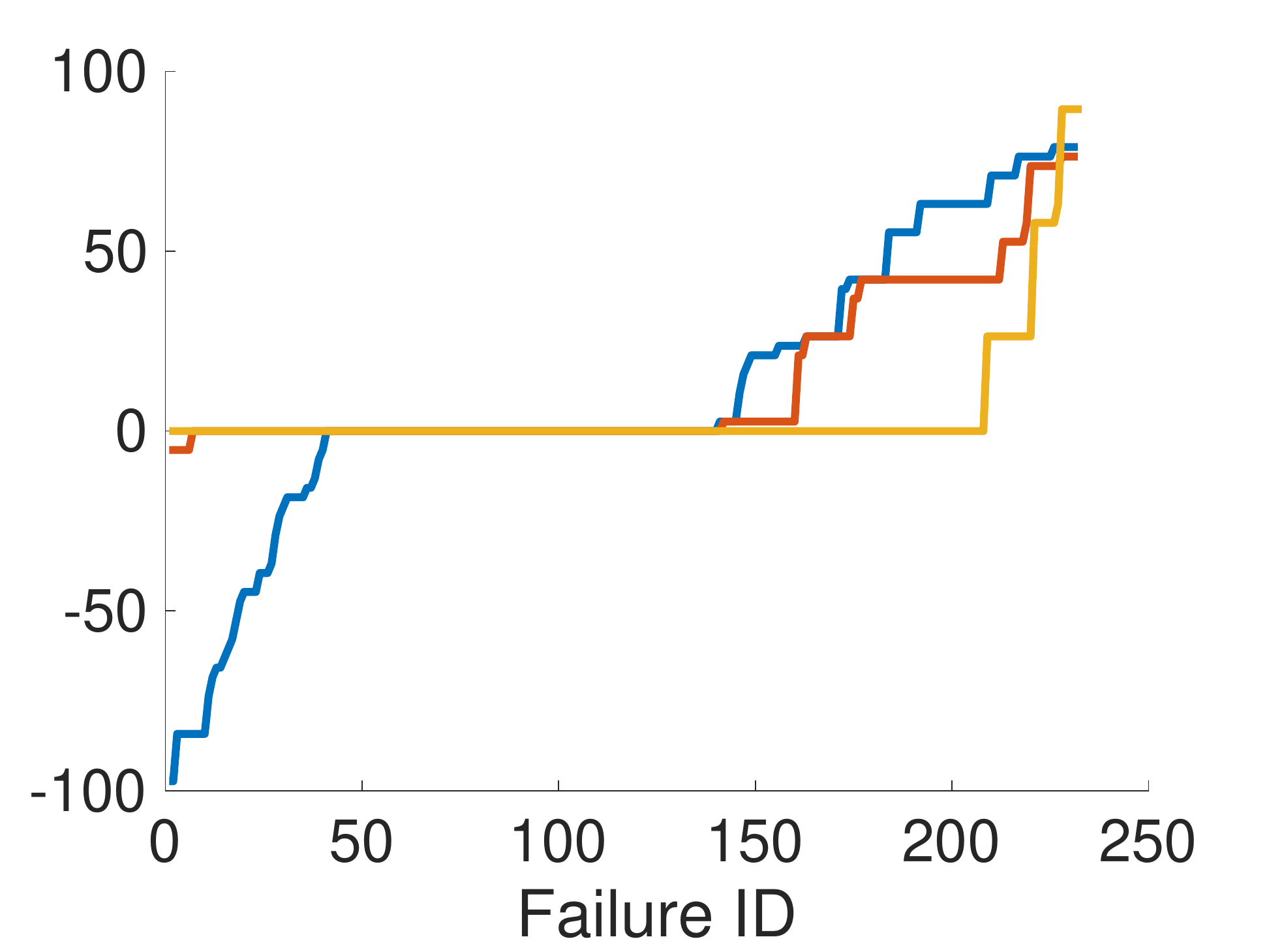}}
\subfloat[240-bus network] {\includegraphics[width=.24\textwidth]{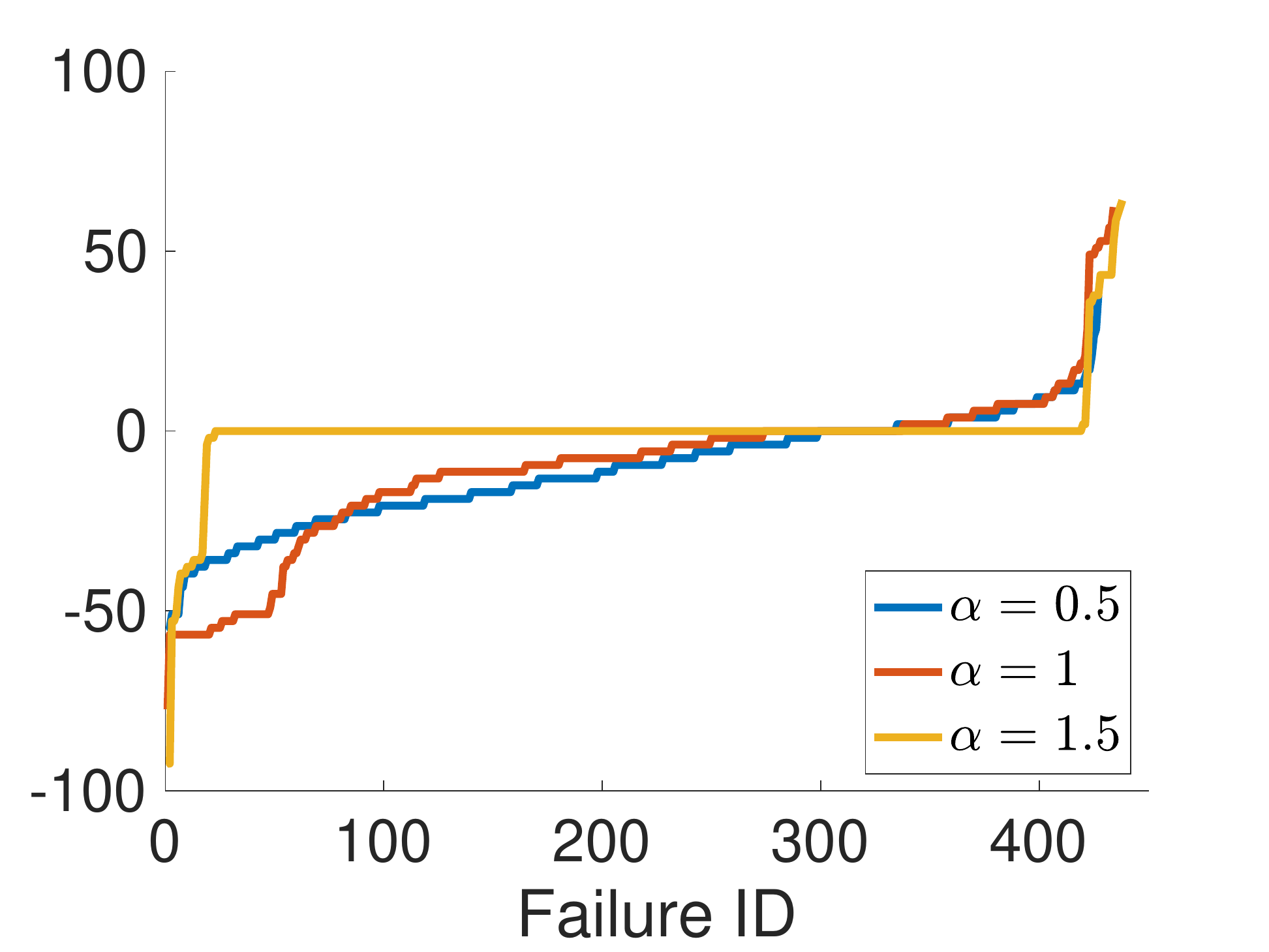}}
\caption{Difference of AGR between tree structure based and mesh structure based approaches.}
\label{fig:tree_mesh}
\end{figure*}
We now demonstrate the failure localization capability of tree partitioning by comparing tree-based and mesh-based approaches (namely UC+T vs. UC+M and AGC+T vs. AGC+M). Similarly, for each pair of approaches, we demonstrate in Figure~\ref{fig:tree_mesh} the difference of AGR for every scenario.

For 118-bus, 179-bus and 240-bus networks, both UC+T and UC+M result in similar amount of failure scenarios with load loss, while the UC+T approach achieves slightly lower LLR on average, as shown in Tables~\ref{table:stat1} and \ref{table:stat2}.  However, in terms of the localization performance, tree-based approaches tend to adjust fewer generators as shown in Figure~\ref{fig:tree_mesh}. More importantly, generators under UC+T always try to mitigate the failure with only local adjustments. Generation adjustments in the control areas without any failures almost never happen.

As in the previous subsection, for the IEEE 200-bus network, UC+T results in more failure scenarios with load loss or adjusted generations than UC+M, possibly due to the sparsity of the original network. Nevertheless, the difference in terms of average load loss is negligible.

\section{Conclusions}\label{section:con}
In this paper we propose an integrated approach to failure mitigation and localization consisting of topology design and real-time response. Both analytical results and case studies validate the potential and capability of our proposed control. There are several research directions for further exploration, including: (i) extending the approach to non-linear system and dynamics, (ii) understanding the optimal trade-off between localization and redundancy, and (iii) managing generation/line reserves to avoid severe failures.
\begin{table*}
\def\arraystretch{1.3}
\centering
\caption{Fraction of failure scenarios with non-zero LLR/AGR over IEEE test networks.} \label{table:stat1}
\begin{tabular}{|c|c|c|c|c|c|c|c|c|c|c|c|c|c|}
\hline
& &\multicolumn{3}{c|}{UC + Tree}&\multicolumn{3}{c|}{UC + Mesh} & \multicolumn{3}{c|}{AGC + Tree} & \multicolumn{3}{c|}{AGC + Mesh}\\
\hline
Network & $\alpha$ & 0.5 & 1 & 1.5 & 0.5 & 1 & 1.5 & 0.5 & 1 & 1.5 & 0.5 & 1 & 1.5 \\
\hline
\multirow{2}{*}{118}& LLR &  48.60\% & 8.24\% & 4.95\% & 43.02\% & 6.04\% & 2.75\% & 96.65\% & 50.00\% & 4.95\% & 98.88\% & 26.37\% & 3.85\%\\
\cline{2-14}
&AGR & 94.41\% & 52.75\% & 8.24\% & 96.09\% & 31.87\% & 6.59\% & 96.65\% & 54.40\% & 8.24\% & 98.88\% & 26.92\% & 6.59\%\\
\hline
\multirow{2}{*}{179}& LLR & 3.88\% & 2.32\% & 1.54\% & 3.88\% & 2.32\% & 1.54\% & 68.22\% & 3.86\% & 2.32\% & 58.53\% & 4.25\% & 1.16\%\\
\cline{2-14}
&AGR & 71.32\% & 44.02\% & 37.07\% & 74.42\% & 69.88\% & 21.24\% & 78.29\% & 38.22\% & 24.71\% & 83.72\% & 63.32\% & 19.69\%\\
\hline
\multirow{2}{*}{200}& LLR & 21.12\% & 18.53\% & 15.02\% & 9.48\% & 7.76\% & 7.30\% & 51.29\% & 37.07\% & 15.02\% & 24.14\% & 7.76\% & 7.30\%\\
\cline{2-14}
&AGR & 56.03\% & 47.41\% & 31.33\% & 36.21\% & 20.26\% & 20.60\% & 56.90\% & 47.41\% & 31.33\% & 36.64\% & 20.26\% & 20.60\%\\
\hline
\multirow{2}{*}{240}& LLR & 22.25\% & 8.53\% & 2.51\% & 24.12\% & 6.91\% & 2.28\% & 96.96\% & 72.81\% & 3.20\% & 97.89\% & 94.70\% & 2.51\%\\
\cline{2-14}
&AGR & 94.61\% & 73.73\% & 21.92\% & 96.72\% & 91.47\% & 16.67\% & 97.66\% & 86.41\% & 18.72\% & 98.13\% & 94.93\% & 15.53\%\\
\hline
\end{tabular}
\end{table*}

\begin{table*}
\def\arraystretch{1.3}
\centering
\caption{Average values of non-zero LLR/AGR over IEEE test networks.} \label{table:stat2}
\begin{tabular}{|c|c|c|c|c|c|c|c|c|c|c|c|c|c|}
\hline
& &\multicolumn{3}{c|}{UC + Tree}&\multicolumn{3}{c|}{UC + Mesh} & \multicolumn{3}{c|}{AGC + Tree} & \multicolumn{3}{c|}{AGC + Mesh}\\
\hline
Network & $\alpha$ & 0.5 & 1 & 1.5 & 0.5 & 1 & 1.5 & 0.5 & 1 & 1.5 & 0.5 & 1 & 1.5 \\
\hline
\multirow{2}{*}{118}& LLR & 1.58\% & 2.34\% & 1.05\% & 2.25\% & 2.65\% & 1.29\% & 17.69\% & 4.22\% & 3.28\% & 18.80\% & 4.64\% & 3.82\%\\
\cline{2-14}
&AGR& 10.37\% & 12.15\% & 15.68\% & 17.68\% & 15.20\% & 18.06\% & 14.56\% & 15.15\% & 16.54\% & 20.52\% & 15.38\% & 17.59\%\\
\hline
\multirow{2}{*}{179}& LLR & 1.59\% & 1.67\% & 1.71\% & 1.56\% & 1.67\% & 1.71\% & 3.33\% & 3.12\% & 2.43\% & 5.41\% & 3.16\% & 2.36\%\\
\cline{2-14}
&AGR & 26.52\% & 26.80\% & 36.71\% & 41.45\% & 42.35\% & 46.39\% & 58.30\% & 62.07\% & 38.79\% & 73.96\% & 75.15\% & 38.67\%\\
\hline
\multirow{2}{*}{200}& LLR  & 2.84\% & 2.07\% & 1.31\% & 2.35\% & 1.86\% & 1.00\% & 17.20\% & 9.94\% & 1.31\% & 13.70\% & 1.86\% & 1.00\%\\
\cline{2-14}
&AGR & 29.86\% & 36.05\% & 50.61\% & 39.94\% & 35.89\% & 50.66\% & 50.42\% & 45.43\% & 50.61\% & 49.57\% & 35.89\% & 50.66\%\\
\hline
\multirow{2}{*}{240}& LLR& 0.93\% & 1.39\% & 2.09\% & 1.00\% & 1.80\% & 2.24\% & 9.36\% & 7.19\% & 2.77\% & 9.78\% & 5.53\% & 3.34\%\\
\cline{2-14}
&AGR & 24.47\% & 21.52\% & 51.93\% & 40.62\% & 45.50\% & 66.24\% & 49.49\% & 50.42\% & 51.61\% & 60.43\% & 56.17\% & 63.12\% \\
\hline
\end{tabular}
\end{table*}
\balance
\bibliographystyle{IEEEtran}
\bibliography{biblio}

\begin{thebibliography}{10}
\providecommand{\url}[1]{#1}
\csname url@samestyle\endcsname
\providecommand{\newblock}{\relax}
\providecommand{\bibinfo}[2]{#2}
\providecommand{\BIBentrySTDinterwordspacing}{\spaceskip=0pt\relax}
\providecommand{\BIBentryALTinterwordstretchfactor}{4}
\providecommand{\BIBentryALTinterwordspacing}{\spaceskip=\fontdimen2\font plus
\BIBentryALTinterwordstretchfactor\fontdimen3\font minus
  \fontdimen4\font\relax}
\providecommand{\BIBforeignlanguage}[2]{{%
\expandafter\ifx\csname l@#1\endcsname\relax
\typeout{** WARNING: IEEEtran.bst: No hyphenation pattern has been}%
\typeout{** loaded for the language `#1'. Using the pattern for}%
\typeout{** the default language instead.}%
\else
\language=\csname l@#1\endcsname
\fi
#2}}
\providecommand{\BIBdecl}{\relax}
\BIBdecl

\bibitem{vaiman2012risk}
M.~Vaiman, K.~Bell, Y.~Chen, B.~Chowdhury, I.~Dobson, P.~Hines, M.~Papic,
  S.~Miller, and P.~Zhang, ``Risk assessment of cascading outages:
  Methodologies and challenges,'' \emph{{IEEE} Transactions on Power Systems},
  vol.~27, no.~2, pp. 631--641, May 2012.

\bibitem{brummitt2003cascade}
C.~D. Brummitt, R.~M. D’Souza, and E.~A. Leicht, ``Suppressing cascades of
  load in interdependent networks,'' \emph{Proceedings of the National Academy
  of Sciences}, vol. 109, no.~12, pp. E680--E689, 2012.

\bibitem{kong2010failure}
Z.~Kong and E.~M. Yeh, ``Resilience to degree-dependent and cascading node
  failures in random geometric networks,'' \emph{{IEEE} Transactions on
  Information Theory}, vol.~56, no.~11, pp. 5533--5546, Nov. 2010.

\bibitem{crucitti2004topological}
P.~Crucitti, V.~Latora, and M.~Marchiori, ``A topological analysis of the
  {I}talian electric power grid,'' \emph{Physica A: Statistical mechanics and
  its applications}, vol. 338, no. 1-2, pp. 92--97, 2004.

\bibitem{hines2017cascading}
P.~Hines, I.~Dobson, and P.~Rezaei, ``Cascading power outages propagate locally
  in an influence graph that is not the actual grid topology,'' \emph{{IEEE}
  Transactions on Power Systems}, vol.~32, no.~2, pp. 958--967, 2017.

\bibitem{carreras2002critical}
B.~A. Carreras, V.~E. Lynch, I.~Dobson, and D.~E. Newman, ``Critical points and
  transitions in an electric power transmission model for cascading failure
  blackouts,'' \emph{Chaos: An interdisciplinary journal of nonlinear science},
  vol.~12, no.~4, pp. 985--994, 2002.

\bibitem{bienstock2010n-k}
D.~Bienstock and A.~Verma, ``The $n-k$ problem in power grids: New models,
  formulations, and numerical experiments,'' \emph{SIAM Journal on
  Optimization}, vol.~20, no.~5, pp. 2352--2380, 2010.

\bibitem{bienstock2011control}
D.~Bienstock, ``Optimal control of cascading power grid failures,'' in
  \emph{2011 IEEE Conference on Decision and Control and European Control
  Conference}, Dec 2011, pp. 2166--2173.

\bibitem{soltan2015analysis}
S.~Soltan, D.~Mazauric, and G.~Zussman, ``Analysis of failures in power
  grids,'' \emph{{IEEE} Transactions on Control of Network Systems}, vol.~4,
  no.~2, pp. 288--300, Jun. 2017.

\bibitem{guo2017critical}
H.~Guo, C.~Zheng, H.~H.-C. Iu, and T.~Fernando, ``A critical review of
  cascading failure analysis and modeling of power system,'' \emph{Renewable
  and Sustainable Energy Reviews}, vol.~80, pp. 9--22, 2017.

\bibitem{bergen2000power}
A.~R. Bergen and V.~Vittal, \emph{Power system analysis}, 2nd~ed.\hskip 1em
  plus 0.5em minus 0.4em\relax Prentice Hall, 2000.

\bibitem{machowski2011power}
J.~Machowski, J.~Bialek, and J.~Bumby, \emph{Power system dynamics: stability
  and control}.\hskip 1em plus 0.5em minus 0.4em\relax John Wiley \& Sons,
  2011.

\bibitem{guo2018failure}
L.~{Guo}, C.~{Liang}, A.~{Zocca}, S.~H. {Low}, and A.~{Wierman}, ``Failure
  localization in power systems via tree partitions,'' in \emph{2018 IEEE
  Conference on Decision and Control (CDC)}, Dec 2018, pp. 6832--6839.

\bibitem{mallada2017optimal}
E.~Mallada, C.~Zhao, and S.~Low, ``Optimal load-side control for frequency
  regulation in smart grids,'' \emph{{IEEE} Transactions on Automatic Control},
  vol.~62, no.~12, pp. 6294--6309, Dec. 2017.

\bibitem{guo2019less}
L.~{Guo}, C.~{Liang}, A.~{Zocca}, S.~H. {Low}, and A.~{Wierman}, ``Less is
  more: Real-time failure localization in power systems,'' in \emph{2019 IEEE
  Conference on Decision and Control (CDC)}, Dec 2019, pp. 3871--3877.

\bibitem{ba2019computing}
Q.~Ba and K.~Savla, ``Computing optimal control of cascading failure in {DC}
  networks,'' \emph{{IEEE} Transactions on Automatic Control}, 2019.

\bibitem{guo2019tps3}
L.~{Guo}, C.~{Liang}, A.~{Zocca}, S.~H. {Low}, and A.~{Wierman}, ``Localization
  \& mitigation of cascading failures in power systems, part iii: Real-time
  mitigation,'' \emph{In preparation}.

\bibitem{zhao2016unified}
C.~Zhao, E.~Mallada, S.~Low, and J.~Bialek, ``A unified framework for frequency
  control and congestion management,'' in \emph{2016 Power Systems Computation
  Conference ({PSCC})}.\hskip 1em plus 0.5em minus 0.4em\relax {IEEE}, Jun.
  2016.

\bibitem{bienstock2007integer}
D.~Bienstock and S.~Mattia, ``Using mixed-integer programming to solve power
  grid blackout problems,'' \emph{Discrete Optimization}, vol.~4, no.~1, pp.
  115--141, 2007.

\bibitem{hines2007controlling}
P.~Hines and S.~Talukdar, ``Controlling cascading failures with cooperative
  autonomous agents,'' \emph{International Journal of Critical
  Infrastructures}, vol.~3, no. 1/2, p. 192, 2007.

\bibitem{albert2004structural}
R.~Albert, I.~Albert, and G.~L. Nakarado, ``Structural vulnerability of the
  north american power grid,'' \emph{Physical review E}, vol.~69, no.~2, p.
  025103, 2004.

\bibitem{zocca2019or}
A.~{Zocca}, L.~{Guo}, C.~{Liang}, S.~H. {Low}, and A.~{Wierman}, ``A spectral
  representation of power systems with applications to adaptive partitioning,
  failure localization, and network optimization,'' \emph{In preparation}.

\bibitem{babaeinejadsarookolaee2019power}
S.~Babaeinejadsarookolaee, A.~Birchfield, R.~D. Christie, C.~Coffrin,
  C.~DeMarco, R.~Diao, M.~Ferris, S.~Fliscounakis, S.~Greene, R.~Huang,
  C.~Josz, R.~Korab, B.~Lesieutre, J.~Maeght, D.~K. Molzahn, T.~J. Overbye,
  P.~Panciatici, B.~Park, J.~Snodgrass, and R.~Zimmerman, ``The power grid
  library for benchmarking {AC} optimal power flow algorithms,''
  \emph{arXiv:1908.02788}, 2019.

\end{thebibliography}

\end{document}